\documentclass[usenatbib, sort]{mn2e}

\usepackage{epsfig}
\usepackage{amssymb}

\title[Cross-Correlation between Damped Ly$\alpha$ Systems and Lyman Break Galaxies]{Cross-Correlation between Damped Ly$\alpha$ Systems and Lyman Break Galaxies in Cosmological SPH Simulations}

\author[Lee et al.]
  {T. S.~Lee$^1$\thanks{Email: tslee@physics.unlv.edu},  
    K.~Nagamine$^1$\thanks{Visiting Researcher, Institute for the Physics and Mathematics of the Universe, University of Tokyo, 5-1-5 Kashiwanoha Kashiwa-shi, Chiba 277-8582 Japan ; Email: kn@physics.unlv.edu}, 
    L.~Hernquist$^2$, 
   and V.~Springel$^3$\thanks{Current address: Heidelberg Institute for Theoretical Studies, gGmbH, Schloss-Wolfsbrunnenweg 35, D-69118 Heidelberg, Germany} 
   \vspace{0.3cm}\\
  $^1$Department of Physics and Astronomy, University of Nevada Las Vegas, 
   4505 Maryland Parkway, Las Vegas, NV 89154-4002, U.S.A.\\
  $^2$Harvard-Smithsonian Center for Astrophysics, 
	60 Garden Street, Cambridge, MA 02138, U.S.A.\\
  $^3$Max-Planck-Institut f\"{u}r Astrophysik, 
	Karl-Schwarzschild-Stra\ss{}e 1, 85740 Garching bei 
	M\"{u}nchen, Germany}

\pagerange{\pageref{firstpage}--\pageref{lastpage}}
\pubyear{2007}

\newcommand{\Mhalo}{M_{\rm halo}}

\def\avg#1{\langle#1\rangle}
\newcommand{\Lam}{\Lambda}

\newcommand{\kms}{\,\rm km\, s^{-1}}
\newcommand{\Msun}{M_{\odot}}

\newcommand{\hinv}{h^{-1}}
\newcommand{\himpc}{\hinv{\rm\,Mpc}}
\newcommand{\hikpc}{\hinv{\rm\,kpc}}
\newcommand{\himsun}{\,\hinv{\Msun}}

\newcommand{\HI}{H\,{\sc i}\,\,}
\newcommand{\NHI}{{N_{\rm HI}}}

\newcommand{\highz}{high-$z$}

\newcommand{\ro}{r_{0}}
\newcommand{\ndla}{n_{\rm DLA}}
\newcommand{\nlbg}{n_{\rm LBG}}
\newcommand{\xDLALBG}{\xi_{\rm DLA-LBG}(r)}
\newcommand{\sDLA}{\sigma_{\rm DLA}}
\newcommand{\rmax}{r_{\rm max}}

\newcommand{\apj}{ApJ}
\newcommand{\apjl}{ApJL}
\newcommand{\mnras}{MNRAS}

\begin{document}

\maketitle

\label{firstpage}


\begin{abstract}

We calculate the cross-correlation function (CCF) between damped
Ly-$\alpha$ systems (DLAs) and Lyman break galaxies (LBGs) using
cosmological hydrodynamic simulations at $z=3$.  We compute the CCF
with two different methods.  First, we assume that there is one DLA in
each dark matter halo if its DLA cross section is non-zero. In our
second approach we weight the pair-count by the DLA cross section of
each halo, yielding a cross-section-weighted CCF.  We also compute the
angular CCF for direct comparison with observations.  Finally, we
calculate the auto-correlation functions of LBGs and DLAs, and their
bias against the dark matter distribution.  For these different
approaches, we consistently find that there is good agreement between
our simulations and observational measurements by \citet{Cooke06a} and
\citet{Ade05f}.  Our results thus confirm that the spatial
distribution of LBGs and DLAs can be well described within the 
framework of the concordance $\Lam$CDM model. 
We find that the correlation strengths of LBGs and DLAs are consistent 
with the actual observations, and in the case of LBGs it is higher 
than would be predicted by low-mass galaxy merger models. 
\end{abstract}

\begin{keywords}
cosmology: theory --- stars: formation --- galaxies: evolution -- galaxies: formation -- methods: numerical
\end{keywords}


\section{Introduction}
\label{section:intro}

According to the cold dark matter (CDM) model of structure formation,
the spatial distribution of galaxies can be understood as a result of
gravitational instability of density fluctuations in the CDM, and
the dark matter halo mass function can be well described by analytic
models \citep{Sheth99}.  More precisely, hierarchical CDM models
predict that the massive galaxies at high redshift (hereafter \highz)
are clustered together in high-density regions, while low-mass
galaxies tend to be more evenly spaced \citep{Kaiser84, Bardeen86}.
Under the assumption that galaxies are produced from primordial
density fluctuations owing to gravitational instability, one can
estimate the average mass of galaxy host haloes based on clustering
data.  For example, \citet{Ade98} estimated the typical halo mass of
LBGs at $z\sim 3$ to be $10^{10.8}\Msun \lesssim \Mhalo \lesssim 10^{12}\,\Msun$ from
observations of their auto-correlation function (ACF).

Damped Lyman-$\alpha$ systems (DLAs), defined as quasar absorption
systems with column density of $\NHI > 2 \times 10^{20}$ atoms
cm$^{-2}$ \citep{Wolfe86}, probe the \HI\ gas associated with
\highz\ galaxies.  Since stars are hardly formed in warm ionized gas
and are tightly correlated with cold neutral clouds, the amount of
\HI\ gas is very important, being the precursor of molecular clouds
\citep{Wolfe03a}.  DLAs dominate the \HI\ content of the Universe at
$z\simeq 3$ and contain a sufficient amount of \HI\ gas mass to
account for a large fraction of the present-day stellar mass
\citep{Storr00}.  The gas kinematics and chemical abundances of DLAs
can be measured and are documented in detail.  However, the masses of
DLA host haloes (hereafter DLA haloes) remain poorly constrained,
because only about $20\%$ of quasars exhibit DLA absorption per unit
redshift \citep{Nag07a}, and the scattered distribution of DLAs in quasar sight
lines precludes the use of DLAs as tracers of dark matter halo mass.

Alternatively, the mass of DLA haloes can be probed by the
cross-correlation between DLAs and a galaxy population whose
clustering and halo mass are well understood.  \citet{Cooke06a,
  Cooke06b} used 211 LBG spectra and 11 DLAs to measure the three
dimensional (3-D) LBG ACF and DLA-LBG CCF \citep[see
  also][]{Gawiser01, Bouche04, Bouche05}.  Their analysis started by
counting the number of LBGs in 3-D cylindrical bins centred on each
of 11 DLAs, following the method of \citet{Ade03}.  They detected a
statistically significant result of DLA-LBG CCF, and estimated an
average DLA halo mass of $\avg{M_{\rm DLA}} \approx
10^{11.2} \Msun$, assuming a single galaxy per halo.

On the theoretical side, \citet{Nag07a} calculated the average DLA
halo mass using a series of cosmological hydrodynamic simulations with
different box sizes, resolution and feedback strengths. They found a
mean DLA halo mass of $\avg{M_{\rm DLA}} = 10^{12.4}\,\Msun$ with their Q5
run which is somewhat larger than $\avg{M_{\rm DLA}} = 10^{11.2}\,\Msun$ of
\citet{Cooke06a}.  
More recent work by \citet{Pontzen08} showed that the DLA
cross-section is predominantly provided by intermediate mass haloes,
$10^9 < M_{\rm vir}/\Msun < 10^{11}$.  These results motivate us to
further examine the distribution of DLAs relative to that of LBGs.  In
this paper, we compute the DLA-LBG CCF in cosmological SPH
simulations, using the sample of LBGs and DLAs obtained by
\citet{Nag04g, Nag04f}.  We compare our results with the observational
results by \citet{Ade05f} and \citet{Cooke06a, Cooke06b}.

Our paper is organized as follows.  In Section~\ref{sec:simulation},
we briefly describe the features of our cosmological SPH simulations
used in this paper.  In Section~\ref{sec:ccf} and
Section~\ref{sec:wccf}, we describe and report the methodology,
binning method, and the results for `unweighted' and `weighted'
DLA-LBG CCF, respectively.  We then discuss the projected angular CCF
for the direct comparison with observational result by
\citet{Cooke06a, Cooke06b} in Section~\ref{sec:angular}.  The ACFs of
LBG-LBG and DLA-DLA are discussed in Subsections~\ref{sec:LBGauto} and
\ref{sec:DLAauto}, while the bias results are reported in
Section~\ref{sec:bias}.  Finally, we discuss the implications of our
work in Section~\ref{sec:discussion}.


\section{Simulations}
\label{sec:simulation}

In this paper, we use two different cosmological smoothed particle
hydrodynamics (SPH) simulations \citep{Springel03a}
performed with the {\small GADGET-2}
code \citep{Springel05e}.  The simulation parameters of the two runs
(named D5 and G5) are summarised in Table~\ref{table:para}.  The same
set of runs has been used by \citet{Nag04f, Nag04g, Nag07a} to study
the global properties of DLAs, such as the DLA cross section,
incidence rate, and \HI\ column density distribution functions.  

The code we use is characterized by four main features. First, it uses
the entropy-conserving formulation of SPH \citep{Springel02}, which
explicitly conserves entropy of the gas where appropriate.  Second,
highly overdense gas particles are treated with a sub-resolution model
for the interstellar medium (ISM) \citep{Springel03b}.  The dense ISM
is assumed to be made of a two-phase fluid consisting of cold
clouds in pressure equilibrium with a hot ambient phase.  Cold clouds
grow by radiative cooling, and form the reservoir of baryons for star
formation. Once star formation occurs, the resulting supernovae (SNe)
deposit energy into the ISM, heating the hot gas environment,
evaporating cold clouds, and transferring cold gas back into the
ambient phase. This establishes a self-regulation cycle for star
formation in the ISM, Additionally, the simulation keeps track of
metal abundance and the dynamical transport of metals.  Metals are
produced by stars and returned into the gas by SNe.

Third, a model for galactic winds is included to study the effects of
outflows on DLAs, galaxies, and intergalactic medium (IGM). In
this model, gas particles are driven out of dense star-forming
medium by assigning an extra momentum in random directions
\citep{Springel03b}.  It is assumed that the wind mass-loss rate is
proportional to the star formation rate, and the wind takes a fixed
fraction of the SN energy.  For the D5 and G5 runs, a strong wind
speed of 484\,km\,s$^{-1}$ is adopted (as opposed to the 'weak' wind speed 
of 242\,km\,s$^{-1}$; \cite{Springel03b}). The dependence of the wind models 
on DLA properties was discussed in detail by \citet{Nag07a}.  Fourth, 
the code includes radiative cooling and heating with a uniform UV 
background of a modified \citet{Haardt96} spectrum \citep{Katz96a, Dave99}, 
where the reionisation takes place at $z\simeq 6$.

We identify simulated galaxies by grouping the star particles using a
simplified variant of the {\small SUBFIND} algorithm proposed by
\citet{Spr01}. This code computes an adaptively smoothed baryonic 
density field for all star and gas particles, and identifies the 
centres of individual galaxies as isolated density peaks.  
It finds the full extent of these galaxies by processing the
gas and star particles in the order of declining density, adding
particles one by one to the galaxies. 

Once the simulated galaxies and consituent particles are identified, 
we then calculate the luminosity and spectrum of individual star 
particles using the mass, formation time, and metallicity using the 
population synthesis code GALAXEV03 \citep{BClib} that assumes the 
Salpeter initial mass function with a mass range of $[0.1, 125]\,\Msun$.
The spectrum of each galaxy is obtained by coadding the spectrum of 
constituent star particles, and the broad-band colours are computed by 
convolving with filter functions. The LBGs are then
selected based on the $U_nGR$ colour selection criteria as described in
\citet{Nag04e}.

\begin{table}
\begin{center}
\begin{tabular}{ccccccc}
\hline
Run & $L_{\rm box}$ & ${N_{\rm p}}$ & $m_{\rm DM}$ & $m_{\rm gas}$ & $\epsilon$ \\
\hline\hline
D5  & 33.75 & $2\times 324^3$ &  $8.15\times 10^7$ & $1.26\times 10^7$ & 4.17  \cr
G5  & 100.0 & $2\times 324^3$ &  $2.12\times 10^9$ & $3.26\times 10^8$ & 12.3  \cr
\hline
\end{tabular}
\caption{Simulations employed in this study.  $N_{\rm P}$ is the
  initial number of gas and dark matter particles (hence
  $\times 2$).  $m_{\rm DM}$ and $m_{\rm gas}$ are the masses of dark
  matter and gas particles in units of $\himsun$, respectively.
  $\epsilon$ is the comoving gravitational softening length in units
  of $\hikpc$, which is a measure of spatial resolution.  All runs
  adopt a `strong' galactic wind feedback model.}
\label{table:para}
\end{center}
\end{table}


\section{Outline of Methods}
\label{sec:method}

Before we present the results of our calculations, let us describe
the series of method that we use to measure the correlations of DLAs and LBGs. 
First, we will examine the simplest case in Section~\ref{sec:ccf}, where we 
assume that there is only one DLA per halo at the center. 
We then discuss the DLA cross-section-weighted CCF in 
Section~\ref{sec:wccf} to examine the effect of multiple DLA clouds in 
massive haloes.  We compute these two cases using the three dimensional 
coordinates ($x,y,z$) of DLAs and LBGs. 
Next, to mimic the observational estimates of two dimensional angular 
CCF, we compute the angular CCF in Section~\ref{sec:angular}.
Finally we compute the auto correlation functions of DLAs and LBGs 
separately in Section~\ref{sec:auto}, and discuss the biases and 
halo masses of DLAs and LBGs in Section~\ref{sec:bias}.


\section{DLA-LBG Cross-Correlation}
\label{sec:ccf}

The probability of finding an object 1 in volume $\delta V_{1}$ at a
separation $r$ from a randomly chosen object 2 can be written as
$\delta P = n_{1}\, [1 + \xi_{12}(r)]\, \delta V_{1}$
\citep{Peebles80book}.  The joint probability of finding an object 1
in volume 1 ($\delta V_{1}$) and an object 2 in volume 2
($\delta V_{2}$) at a separation $r$ is defined as $\delta P = n_{1}
n_{2}\, [1 + \xi_{12}(r)]\, \delta V_{1} \delta V_{2},$ where $n_{1}$
and $n_{2}$ are the mean number densities of the two population.  For
the cross-correlation, we replace object 1 and 2 with DLA and LBG,
then the joint probability between DLA and LBG is
\begin{eqnarray}
\delta P = \ndla \nlbg\, [1 + \xDLALBG]\, \delta V_{\rm DLA} \delta V_{\rm LBG},
\label{eq:proba}
\end{eqnarray}
where $\ndla$ and $\nlbg$ are the mean number densities of DLAs and LBGs, 
and $\xDLALBG$ is the cross-correlation function 
(CCF).

To estimate $\xDLALBG$, we use the method of \citet{Landy93} and \citet{Cooke06a}:
\setlength\arraycolsep{1pt}
\begin{eqnarray}
\lefteqn {\xDLALBG = } \nonumber\\
&&\frac{D_{\rm DLA}D_{\rm LBG}-D_{\rm DLA}R_{\rm LBG}-R_{\rm
    DLA}D_{\rm LBG}+R_{\rm DLA}R_{\rm LBG}}{R_{\rm DLA}R_{\rm LBG}},\nonumber\\
\label{eq:xi01}
\end{eqnarray}
where $D_{\rm DLA}D_{\rm LBG}$ is the number of pairs between the two
data samples of DLAs and LBGs separated by a distance $r \pm \delta
r$, and likewise for other terms.  The notation ``$R_{\rm DLA}$'', for
example, represents the DLA sample that has random coordinate
positions but with an equivalent number density as the original data
sample ``$D_{\rm DLA}$''.

\begin{figure*}
\begin{center}
\resizebox{8.5cm}{!}{\includegraphics{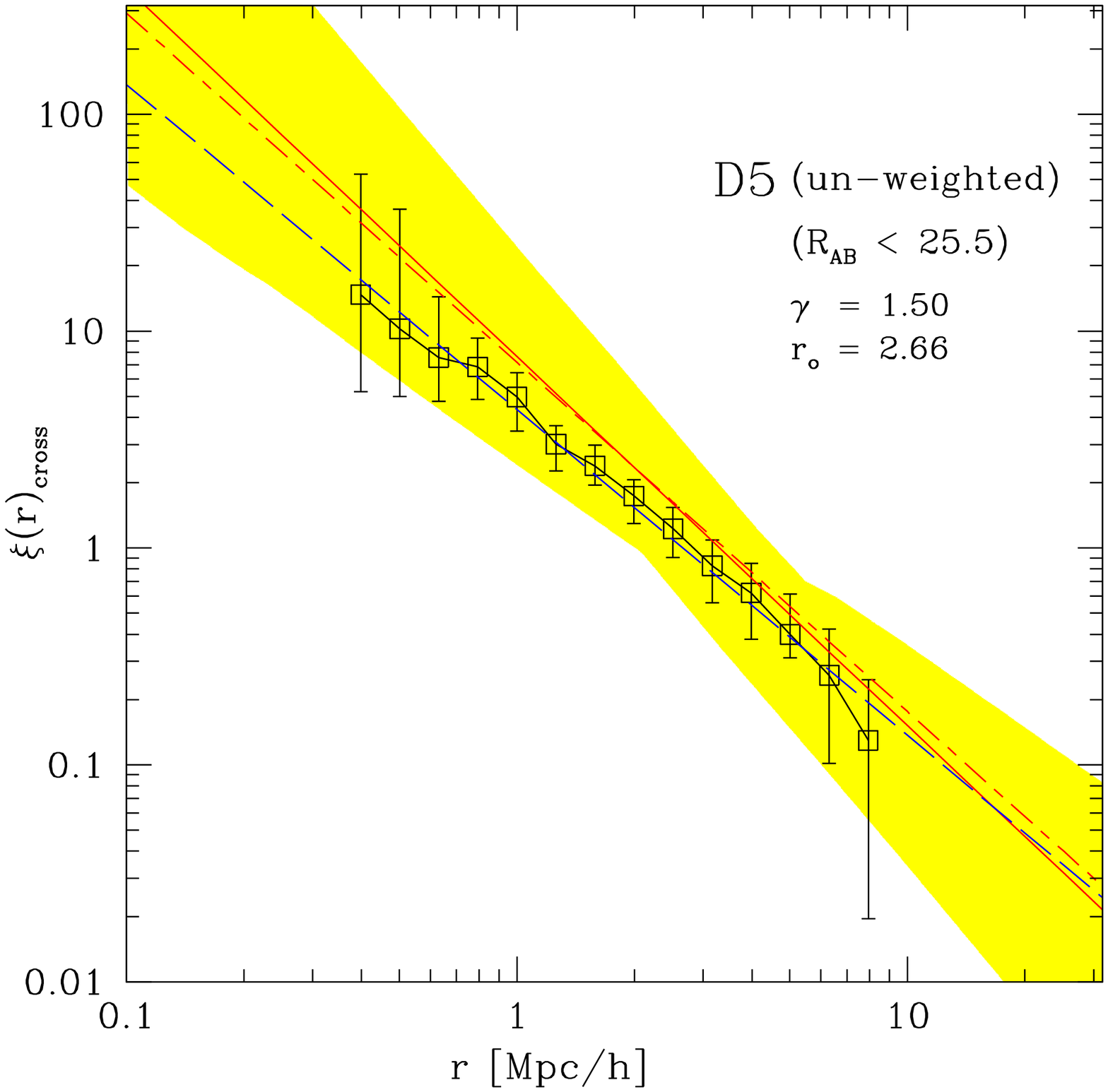}}
\hspace{0.3cm}
\resizebox{8.5cm}{!}{\includegraphics{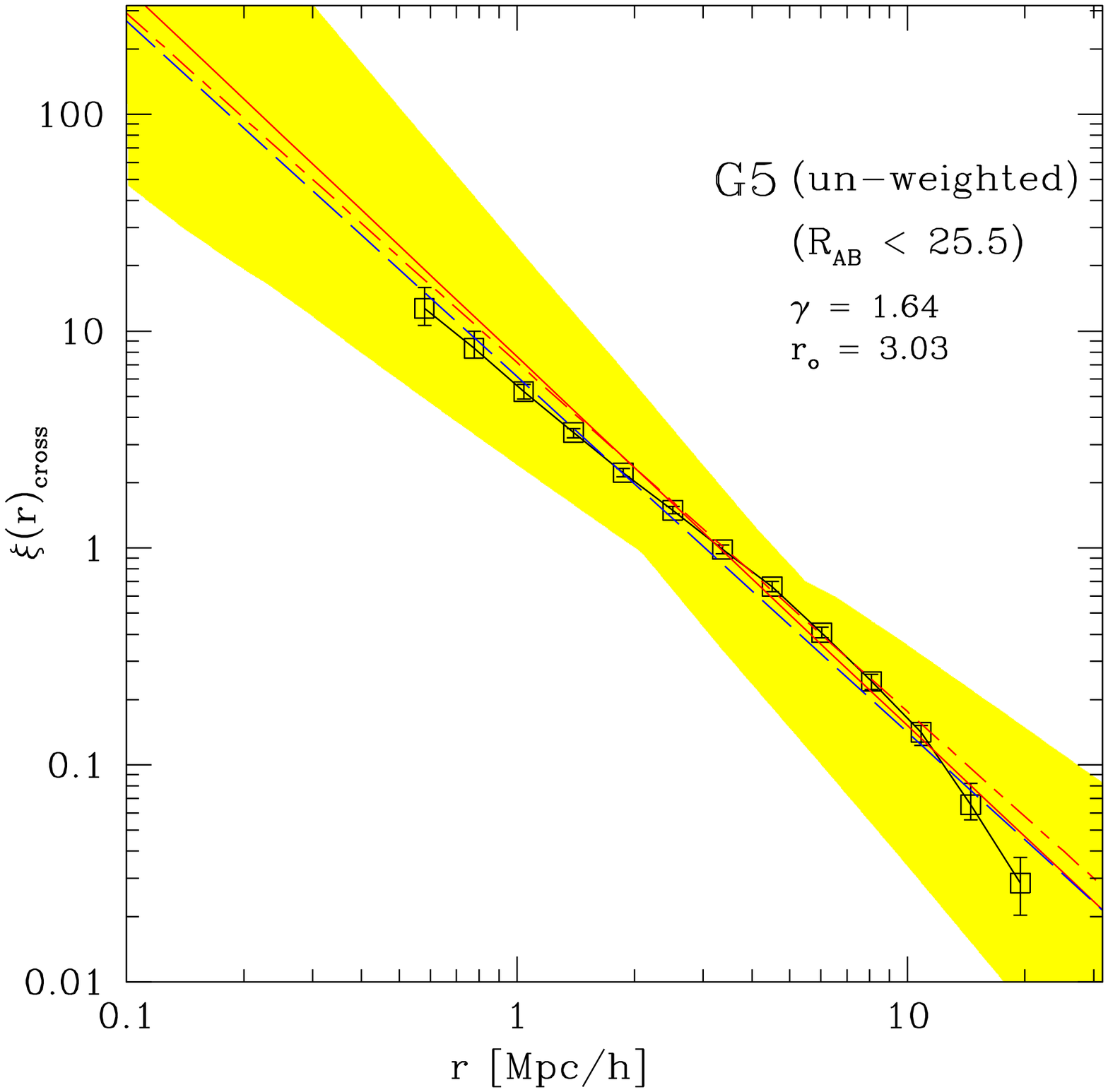}}
\caption{DLA-LBG CCFs at $z=3$ calculated with the regular {\it
    unweighted} method (Equation~\ref{eq:xi01}).  The variance of CCFs
  computed with 100 different random seeds is shown with vertical errorbars, 
  and the open square symbols are the mean of 100 trials.  The
  blue dashed line is the least-square fit to the open square points.
  The red solid line and the short and long dashed lines are the
  angular and 3-D best-fitting power-laws of \citet{Cooke06a,
    Cooke06b}, respectively, and the yellow shade is their 1-$\sigma$
  error range for the angular CCF. }
\label{fig:ccuw}
\end{center}
\end{figure*}

The method of identifying DLAs in our simulations is described in
detail in \citet{Nag04g} \citep[see also][]{Katz96b, Hern96}.
Briefly, we set up a cubic grid that
completely covers each dark matter halo, with the grid-cell size
equivalent to the gravitational softening length `$\epsilon$' of each
run. We then calculate the \HI\ column density $\NHI$ of each pixel
(i.e., a grid-cell on one of the planes) by projecting the \HI\ mass
distribution, and identify those that exceed the DLA threshold of
$\NHI > 2\times 10^{20}$ atoms cm$^{-2}$ as `DLA-pixels'.  This method
allows us to quantify the DLA cross-section `$\sDLA$' of each halo,
and the number of DLA-pixel is $N_i^{\rm DLA} = \sDLA / \epsilon^2$.
Note that we have not made any corrections for the self-shielding of neutral gas in this work. Self-shielding may have significant impact at high \HI\ column densities, and we are currently investigating this issue with separate simulations \citep{Nag10}.

Here we focus on the correlation signal at $r \gtrsim
0.4\,\himpc$, because this is the scale probed by \citet{Cooke06a,
  Cooke06b}. Therefore in this paper we are only concerned about the
overall halo positions and not the exact locations of individual
DLA-pixel within each halo.  The $\sDLA$-weighted CCF will be
discussed in Section~\ref{sec:wccf}.

First, we select the LBGs that are brighter than $R_{\rm AB}$=25.5
magnitude in the D5 and G5 runs.  There are 30 (4030) LBGs in the D5
(G5) run.  \citet{Nag04e} have shown that the brightest galaxies with
$R_{\rm AB} < 25.5$ in our simulations satisfy the $U_nGR$ colour
selection criteria for LBGs \citep[e.g.,][]{Steidel99}. Figure 2 of 
\citet{Nag04e} shows that the D5 run underestimates the number density of
LBGs, while the G5 run agrees better with the observation.

There are 22616 (25683) DLA haloes with $\sDLA > 0$ in the simulated volumes of
the D5 (G5) run.  The `random' catalogues of LBGs and DLA haloes with
random positions were created with a random number generator from
Numerical Recipes \citep{Press92book}.  The selected LBGs were paired
with DLA haloes, and the number of pairs that reside in spherical
shells of [$\log r, \log r + \Delta \log r$] were counted.  The maximum
pair separations probed for the D5 and G5 runs are 10 and
$35\,\himpc$, respectively, with 20 bins in a logarithmic scale of
distance $r$.  The periodic boundary condition was taken into account
appropriately, and the pair-search was extended to the next adjacent
box where needed.

We correct all $\xi(r)$ values by the integral constraint (IC).
This correction owes to the finite size of the
observed field-of-view, and it must be added to the computed
correlation function as follows:
\begin{table}
\begin{center}
\begin{tabular}{ccccccc}
\hline
& \multicolumn{2}{|c|}{Unweighted} & \multicolumn{2}{|c|}{$\sDLA$-weighted} \\

Run &  $\ro$ & $\gamma$ & $\ro$ & $\gamma$\\
\hline\hline
D5 &  2.66 $\pm$ 0.23 & 1.50 $\pm$ 0.17 & 3.37 $\pm$ 0.36 & 1.77 $\pm$ 0.21 \cr
G5 &  3.03 $\pm$ 0.04 & 1.64 $\pm$ 0.03 & 3.43 $\pm$ 0.06 & 1.66 $\pm$ 0.03 \cr
\hline
\end{tabular}
\caption{Best-fitting power-law parameters of unweighted and
  $\sDLA$-weighted DLA-LBG CCFs at $z=3$.  The correlation length
  $r_0$ is in units of $\himpc$.  The confidence limit statistics for this work
  are described in Section~\ref{sec:chisq}.  For comparison, Cooke (private
  communication) obtained $\ro = 2.91 \pm 1.0 \,\himpc$ and $\gamma
  = 1.21^{+0.6}_{-0.3}$ for the 3-D CCF calculated with spherical
  shells, and \citet{Cooke06b} reported $\ro = 3.32\pm
  1.25\,\himpc$ and $\gamma = 1.74\pm 0.36$ for the {\it angular}
  CCF.}
\label{table:ccfit}
\end{center}
\end{table}
\begin{eqnarray}
\xi'(r) = \xi(r) + {\rm IC},
\label{eq:xicorrect}
\end{eqnarray}
where $\xi'(r)$ and $\xi(r)$ are the corrected and computed CCF (or ACF)
respectively.  Following the method
described in \citet{Ade05f} and \citet{LeeKS06}, we calculate the
value of IC and find that it changes $\xi(r)$ only slightly
in our simulations, with IC $\sim 10^{-2}$ for the D5 run.

Figure~\ref{fig:ccuw} shows the DLA-LBG CCF computed with
Eq.\,(\ref{eq:xi01}).  We perform a least-square fit to the measured
values with a power-law $\xi(r)=(\ro / r)^{\gamma}$, and find
best-fitting parameters equal to $(\ro [\himpc], \gamma) = (2.66\pm0.23,
1.50\pm0.17)$ and $(3.03\pm0.04, 1.64\pm0.03)$ for the D5 and G5 runs, 
respectively. The fits are shown by the blue long-dashed lines (see also
Table~\ref{table:ccfit}), and the confidence limit statistics are described 
in Section~\ref{sec:chisq}.  

\citet{Landy93} showed that the variance of $\omega_p (r_{\theta})$ 
obtained from Monte Carlo calculations agrees quite 
well with the standard Poisson variance. Here, we
follow their method outlined in their Section 5.2 and repeat the calculation 
of the CCF 100 times using different seeds for generating the random positions 
for the `random' sample to examine the statistical variance of the
measured CCF. The variance of 100 trials is shown as vertical errorbars,
and the average of 100 trials is shown with the open square data
points.  The red solid line and the yellow shade represent the
best-fitting result ($\ro = 3.32\pm 1.25$ and $\gamma = 1.74\pm
0.36$) and the 1-$\sigma$ errors of \citet{Cooke06a, Cooke06b} from
their angular CCF result.  The result of the G5 run agrees well
with that of Cooke et al.'s, and its variance is small owing to a
larger sample than in the D5 run.  The result of the D5 run is
somewhat shallower than that of the G5 run, which could simply owe
to relatively small sample of LBGs in D5 and its small box-size.

\citet{Cooke06a, Cooke06b} published only the angular CCFs. However,
they can also estimate the 3-D radial CCF using redshift
information.  The best-fitting parameters to the radial CCF by Cooke
(private communication) using spherical shells is $\ro = 3.39 \pm
1.2 \,\himpc$ and $\gamma = 1.61 \pm 0.3$, which is shallower than the
angular CCF results.  As we will further discuss in
Section~\ref{sec:angular}, the method of \citet{Ade03} adopts
cylindrical shells at small distances, which have larger volumes than
spherical shells.  The cylindrical shell method uses long cylinders at small 
$r_{\theta}$ and captures all the potential LBGs near the DLAs, whereas the 
spherical bins do not. This effect seems to result in the slightly steeper 
$\gamma$ in \citet{Cooke06b} compared to the above
spherical shell case (Cooke; private communication).  We regard the comparison 
to the angular CCF of \citet{Cooke06b} as the primary one, because Cooke 
et al. argue that the angular CCF calculated by the method of \citet{Ade03} 
is more robust than the 3-D radial calculation with spherical shells, and the
values of ($\ro$, $\gamma$) derived from both CCFs should be equivalent 
theoretically (see Section~\ref{sec:angular}).

\begin{figure*}
\begin{center}
\resizebox{8.2cm}{!}{\includegraphics{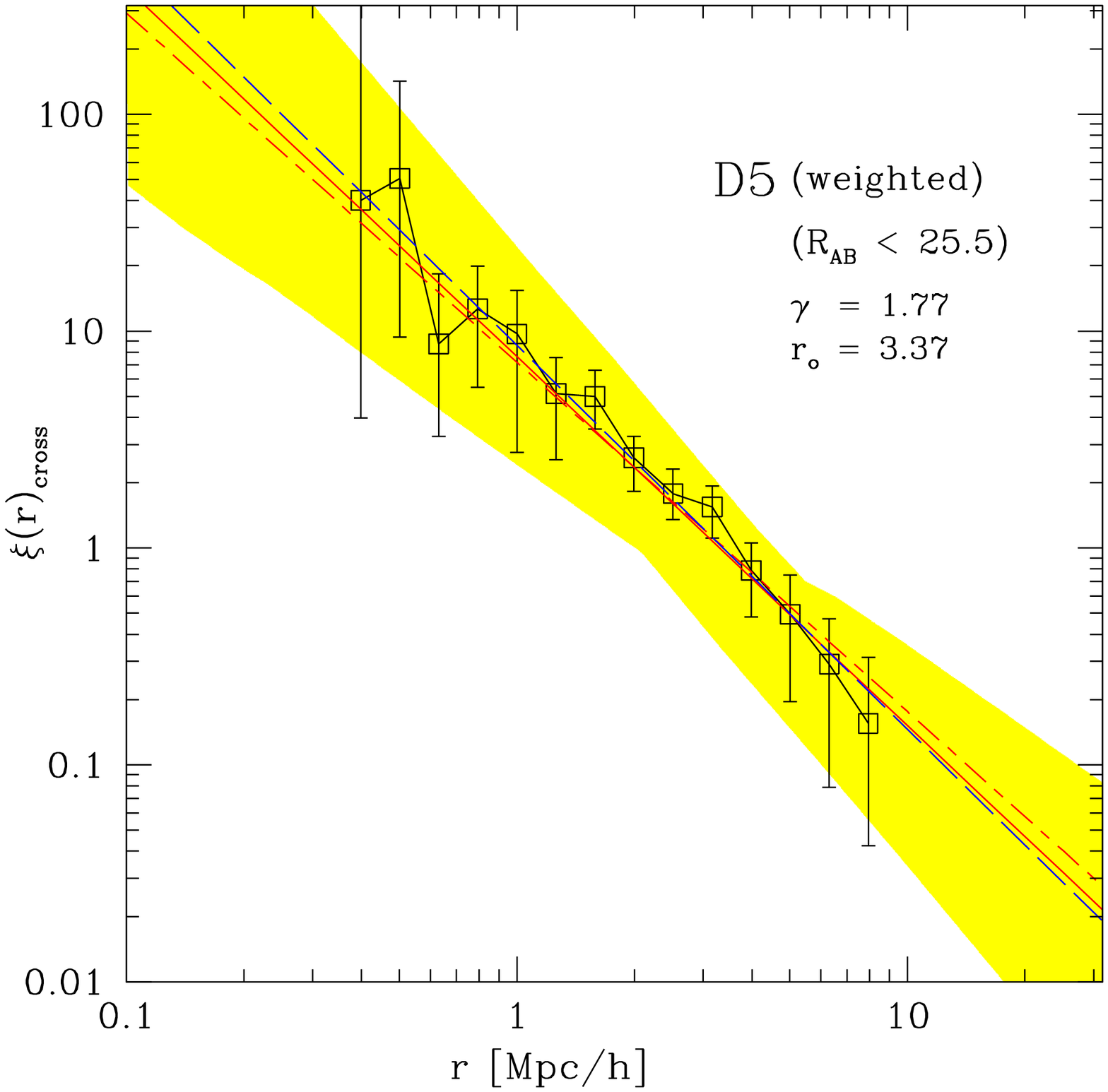}}
\hspace{0.3cm}
\resizebox{8.2cm}{!}{\includegraphics{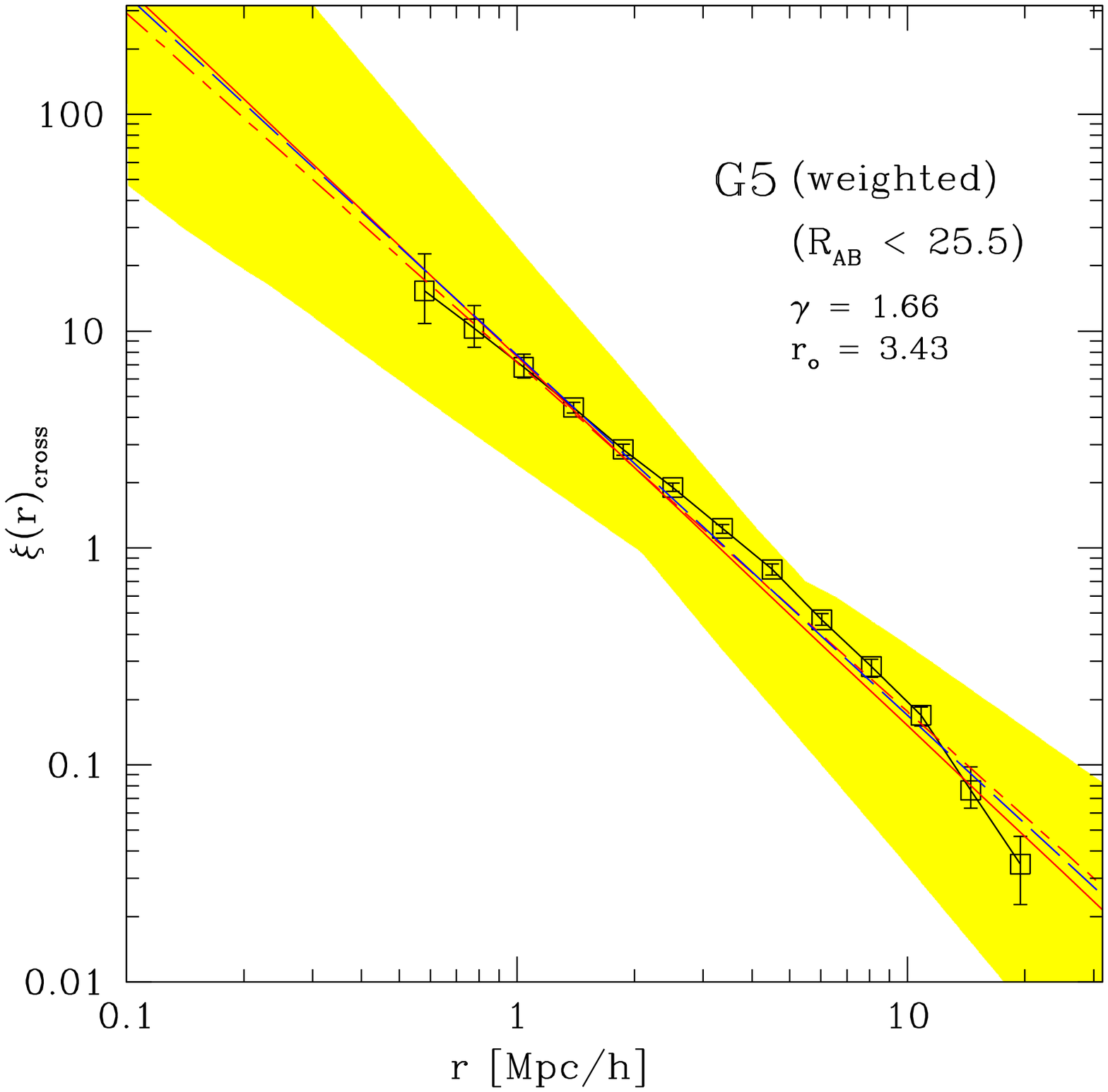}}
\caption{DLA-LBG CCF at $z=3$ calculated by the $\sDLA$-weighted
  method (Equation~\ref{eq:xi02}).  The yellow shade shows the upper
  and lower limits of \citet{Cooke06a, Cooke06b}'s best-fitting
  power-laws. The variance of the CCF using 100 random seeds shown with 
  vertical errorbars. The blue dashed lines are the best-fittings for
  this work, and the red solid line and the short and long dashed
  lines are, respectively, the angular and 3-D best-fitting power-laws
  of \citet{Cooke06a, Cooke06b}.}
\label{fig:ccw}
\end{center}
\end{figure*}


\section{$\sDLA$-weighted CCF}
\label{sec:wccf}

In Section~\ref{sec:ccf}, we calculated the CCF assuming that there is
one DLA per halo.  This assumption is valid as long as we are
concerned with the CCF at scales of $r \gtrsim 300\,\hikpc$. However,
\citet[][Fig.~1]{Nag04f} showed that the DLA clouds have extended distributions 
in massive dark matter haloes.  Therefore, it may be more desirable to take the DLA
cross-section of each halo into account when calculating the CCF, because the chance
of finding a DLA in the actual observation is already cross-section weighted. Ideally,
we would use all the DLA pixels and find pairs with the nearby LBGs, but this 
computation is prohibitively expensive owing to the large number of DLA pixels.

A simple way to achieve this is to weight the number of DLA-LBG pairs by
the number of DLA-pixels of each halo.  Since the displacement between
DLA-pixels within a single halo is typically much smaller than the
distance between LBG-DLA pairs, we do not count the individual pairs
between LBG and DLA-pixels. Instead, we treat it as if all DLA-pixels are
located at the halo centre, and weight each DLA-LBG pair-count by the
number of DLA-pixels $N_i$ (hereafter we drop the superscript `DLA'
for simplicity) and compute the $\sDLA$-weighted CCF as
\setlength\arraycolsep{2pt}
\begin{eqnarray}
\lefteqn{\xDLALBG =  {} } \nonumber\\
& & {} \frac{\scriptstyle N_{i}D_{\rm DLA}D_{\rm LBG}-N_{i}D_{\rm
    DLA}R_{\rm LBG}-N_{i}R_{\rm DLA}D_{\rm LBG}+N_{i}R_{\rm DLA}R_{\rm
    LBG}}{\scriptstyle N_{i}R_{\rm DLA}R_{\rm  LBG}} {}. \nonumber\\
\label{eq:xi02}
\end{eqnarray}
For the `random' DLA dataset, we shuffle the original $N_{i}$ list
randomly and make new pairs with different DLA haloes.  Again, 10
realisations of the random dataset have been used to examine
the statistical variance of the estimated CCF.

The results for the $\sDLA$-weighted CCF is shown in Figure~\ref{fig:ccw}.  
We find best-fitting parameters of 
($\ro\,[\himpc], \gamma) = (3.37\pm0.36, 1.77\pm0.21)$ and 
$(3.43\pm0.06, 1.66\pm0.03)$ for the D5 and G5 runs, respectively, 
as shown by the blue long-dashed line (see also Table~\ref{table:ccfit}).  
(See Section~\ref{sec:chisq} for the error estimates.)
Both results show good agreement with the best-fitting values of 
\citet[][$\ro = 3.32\pm1.25$ and $\gamma = 1.74\pm 0.36$]{Cooke06b}.  
The result of D5 is somewhat noisy at $r\lesssim 1\,\himpc$, which 
originates from the noisy pair-count of $N_i D_{\rm DLA} D_{\rm LBG}$.

The parameter values given in Table~\ref{table:ccfit} clearly show
that the $\sDLA$-weighted method gives larger values of $\ro$ and a
slightly steeper power-law slope.  In a CDM universe, the number of
low-mass haloes is far greater than that of massive haloes.  Therefore,
even a small weighting by $N_i$ boosts up the overall pair-count,
yielding a stronger correlation signal compared to the unweighted
case. The larger LBG sample in the G5 run makes its result more robust
against the weighting procedure than that of the D5 run. Therefore, the 
difference in the slope $\gamma$ between the two calculation methods is 
smaller in the G5 run than that of D5 run.

\subsection{Confidence Limits}
\label{sec:chisq}

The $\chi^{2}$ test describes the goodness-of-fit of the model to the data.  
To determine the confidence intervals of the two parameters ($\gamma$ and $\ro$), 
we use the minimum $\chi^{2}$ method. This statistic is written as
\begin{eqnarray}
\chi^{2} \equiv \sum_{i=1}^{n} \frac{(O_{i}-E_{i})^{2}}{\sigma_{i}^{2}},
\end{eqnarray}
where $O_{i}$ are the data points shown in the correlation figures, 
$E_{i}$ are the expected values in each bin $i$ from the power-law, 
and $\sigma_{i}$ is the standard deviations in each bin obtained from the 
100 Monte Carlo calculations, as described earlier.  

\begin{figure*}
\begin{center}
\resizebox{8.2cm}{!}{\includegraphics{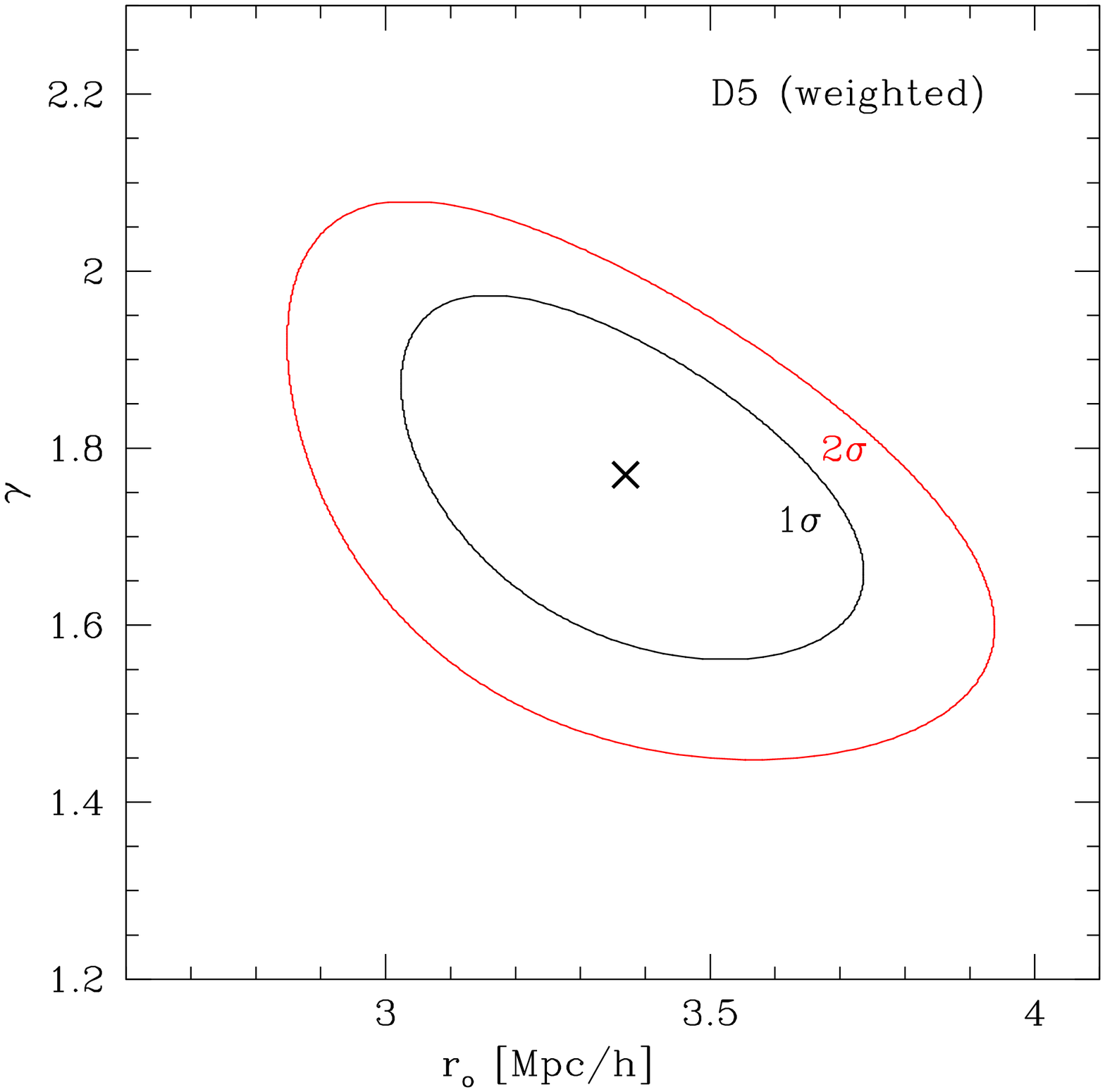}}
\hspace{0.3cm}
\resizebox{8.2cm}{!}{\includegraphics{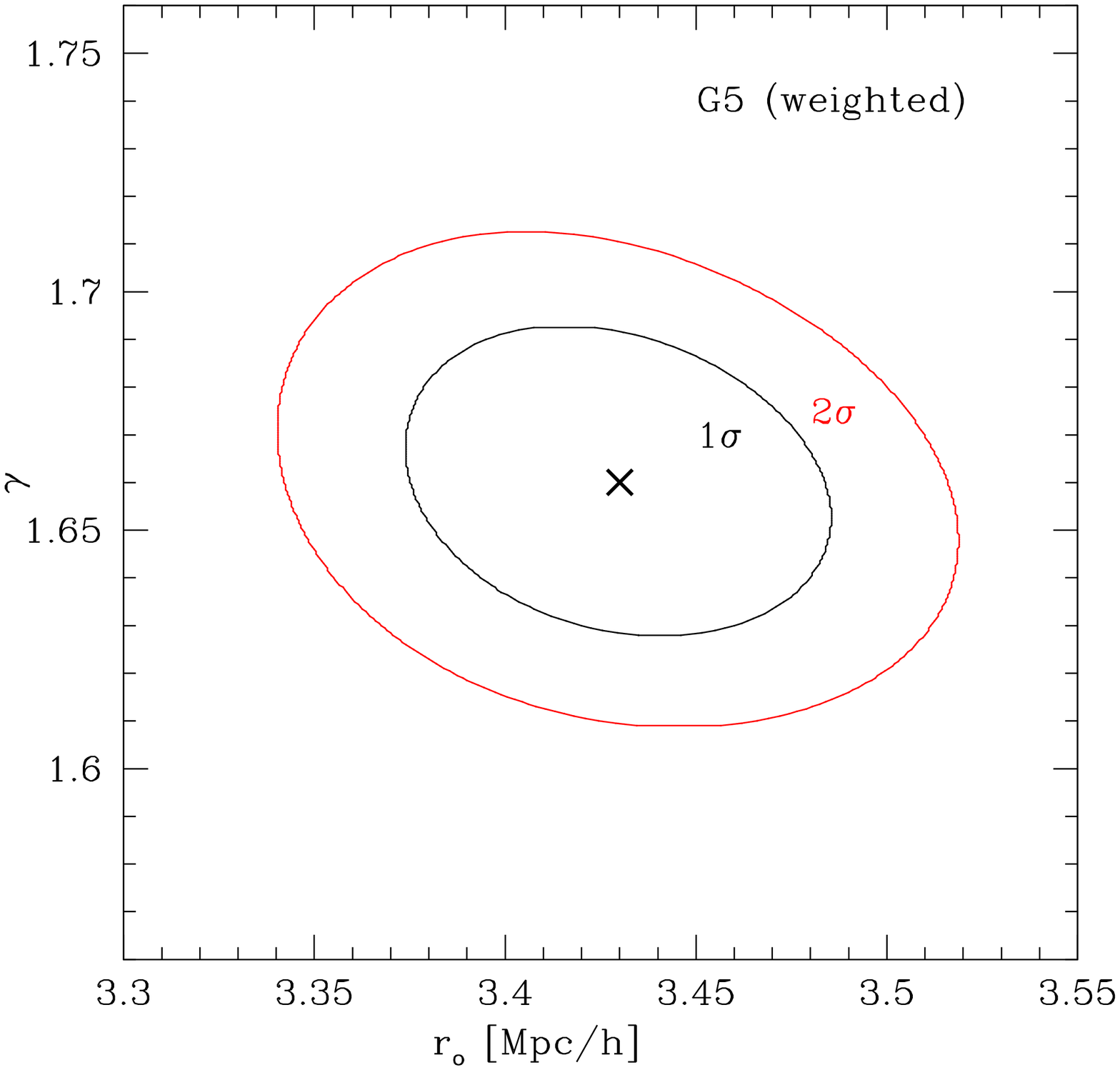}}
\caption{Two parameter confidence limit contours for the weighted DLA-LBG cross
correlation case using the minimum $\chi^{2}$ method.  The best fits of the two parameters are indicated by the cross at the center of the contours, and 1 and 2-$\sigma$ limits are shown in black and red contour lines, respectively.}
\label{fig:chisq}
\end{center}
\end{figure*}

The region of confidence limits \citep{Avni76} is given by 
\begin{eqnarray}
\label{eq:chisq02}
\chi^{2}_{p} = \chi^{2}_{\rm min} + \Delta(df,p),
\end{eqnarray}
where $p$ is a confidence level ($0 < p < 1$), and $df$ is a degree of freedom written as $df = n - c$, where $n$ is the number of bins and $c$ is the number of parameters. For this work, $c = 2$ and $n= 13$ for G5 DLA-LBG CCFs (un-weighted and $\sigma_{DLA}$-weighted), G5 LBG-LBG ACF, and G5 DLA-DLA ACFs (un-weighted and $\sigma_{DLA}$-weighted); $n= 14$ for D5 DLA-LBG 3D and $angular$ CCFs (un-weighted and $\sigma_{DLA}$-weighted); and $n=17$ for G5 DLA-LBG $angular$ CCFs (un-weighted and $\sigma_{DLA}$-weighted).  
The value $\Delta(df,p)$ is the expected increment of $\chi^{2}$ to find the 68$\%$ and 95$\%$ confidence limits above $\chi^{2}_{\rm min}$.  Its value is determined by the degree of freedom and probability within 1 and 2-$\sigma$ limits.  We calculate the 1-$\sigma$ confidence limits for all the correlation cases using this method.  As an example, we show the 1 and 2-$\sigma$ confidence levels for the weighted CCFs of D5 and G5 runs in Figure~\ref{fig:chisq}.

\section{Angular cross correlation function}
\label{sec:angular}

In observational studies, a different method is usually used
to obtain the values of
($r_0$, $\gamma$) compared with what we described in
Sections~\ref{sec:ccf} and \ref{sec:wccf}, because the
precise estimation of any LBG position along the line of sight is
difficult to achieve owing to redshift uncertainties caused by peculiar
velocities and galactic winds.  With such imprecision, it is not
possible to measure the CCF at scales $r\lesssim 1\himpc$ reliably.
Therefore, rather than attempting to estimate the 3-D distance between
DLAs and LBGs, observers usually employ the angular CCF using the
projected data on the sky.  For example, \citet{Cooke06a, Cooke06b}
computed the angular CCF using the method proposed by \citet{Ade03}.
In order to compare our results with those by Cooke et al's, we
briefly describe the calculation method of \citet{Ade03}, and then
describe how we perform our measurement of the angular CCF.

With a power-law assumption, the expected number of pairs for the projected 
angular CCF is
\setlength\arraycolsep{2pt}
\begin{eqnarray}
\label{eq:apccf01}
\omega_p (r_{\theta} < r_z) = 
\frac{\ro^{\gamma} r_{\theta}^{1-\gamma}}{2r_z} B\left( \frac{1}{2},\frac{\gamma-1}{2} \right) I_x \left( \frac{1}{2},\frac{\gamma-1}{2} \right),
\end{eqnarray}
where $B$ and $I_x$ are the beta and incomplete beta functions with \citep[e.g.,][]{Press92book}
\setlength\arraycolsep{2pt}
\begin{eqnarray}
\label{eq:xztheta}
x \equiv r_z^2 \left(r_z^2 + r_{\theta}^2 \right)^{-1}.
\end{eqnarray}
\citet{Ade03} proposed to count the number of pairs in cylindrical shells of 
angular separation $r_{\theta} \pm \delta r_{\theta}$ 
and redshift separation $r_z \pm \delta r_z$, rather than using 
spherical shells.  By setting $r_z$ to 
\setlength\arraycolsep{2pt}
\begin{eqnarray}
\label{eq:zmax}
r_{z}={\rm max}\left(1000\,{\kms} \frac{(1+z{x})}{H(z)}, 7r_{\theta} \right),
\end{eqnarray}
the lower limit ensures that the redshift errors do not lead to the underestimate of 
the number of pairs, and the upper limit allows sufficient distances to 
include enough correlated pairs \citep{Ade03}. 

For our calculations, we focus at $z=3$ and thus 
$r_z = {\rm max}(12.8\,\himpc, 7r_{\theta})$.  With simple algebra, 
Equation~(\ref{eq:apccf01}) can be converted to a more familiar power-law form:
\setlength\arraycolsep{2pt}
\begin{eqnarray}
\label{eq:apccf02}
\xi(r_{\theta}) &=& 2 \rmax \frac{\omega_p (r_{\theta})}{r_{\theta}} \biggl[ B \left( \frac{1}{2},\frac{\gamma-1}{2} \right) I_{x} \left( \frac{1}{2}, \frac{\gamma-1}{2} \right) \biggr]^{-1} \nonumber \\
&=& \left( \frac{r_{\theta}}{\ro} \right)^{-\gamma},
\end{eqnarray}
where $r_z$ is set to $\rmax$.  We change from spherical coordinates
to cylindrical coordinates, and set the number of cylindrical bins to
20 in a logarithmic scale as before.  All pair searches are extended
to the adjacent box using periodic boundary conditions, if appropriate.

A few assumptions must be made while we deal with the beta and
incomplete beta functions. There are two parameters ($\gamma$ and $x$)
that must be given to calculate the values of $B$ and $I_{x}$.  To
calculate $\gamma$, we first plot Equation~(\ref{eq:apccf02}) without
$B$ and $I_x$ (i.e., $2\rmax \omega_p (r_{\theta}) / r_{\theta}$) and
find the best-fitting value of $\gamma$.  The value of $x$ is
determined by $r_z$ and $r_{\theta}$ as shown in
Equation~(\ref{eq:xztheta}).  By setting $r_z = \rmax$, the angular
separation will be divided into two different regimes. Within the
smaller angular separation range ($100\,\hikpc < r_{\theta} <
1.83\,\himpc$), the correlated pairs are counted up to the maximum
radial distance of $\rmax = \pm 12.8\,\himpc$ for a cylinder centred
on an LBG or DLA, while in the larger separation range ($r_{\theta} >
1.83\,\himpc$) all the correlated pairs within $\rmax = \pm 7
r_{\theta}$ are counted. We calculate the values of $B$ and $I_{x}$ (as well
as the IC correction) separately for the two different $r_{\theta}$ regions.
With the fixed values of $\gamma$ obtained above and 20 different values of 
$x$, $B$ and $I_{x}$ can be calculated for each bin.

\begin{figure*}
\begin{center}
\resizebox{8.2cm}{!}{\includegraphics{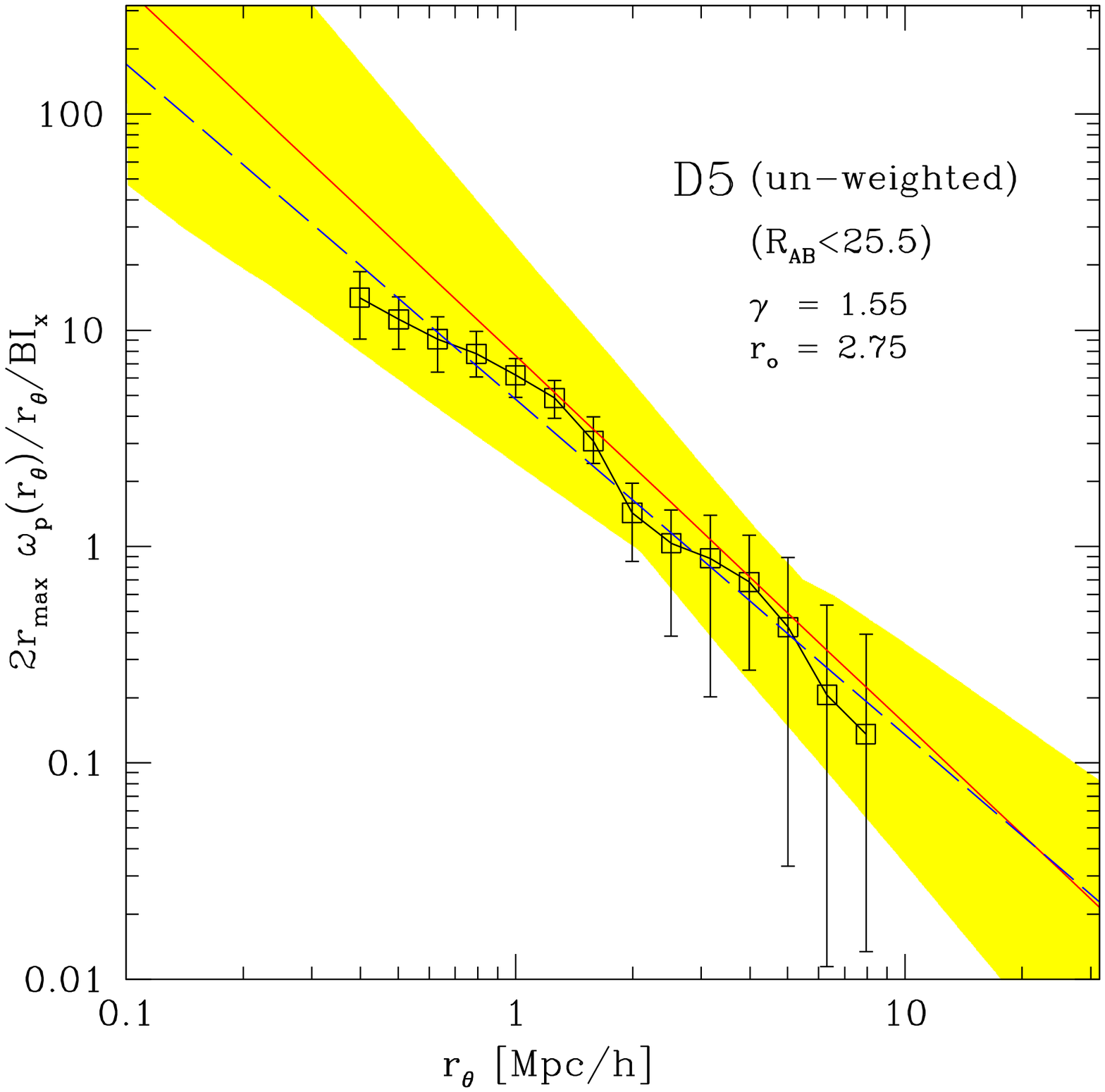}}
\hspace{0.3cm}
\resizebox{8.2cm}{!}{\includegraphics{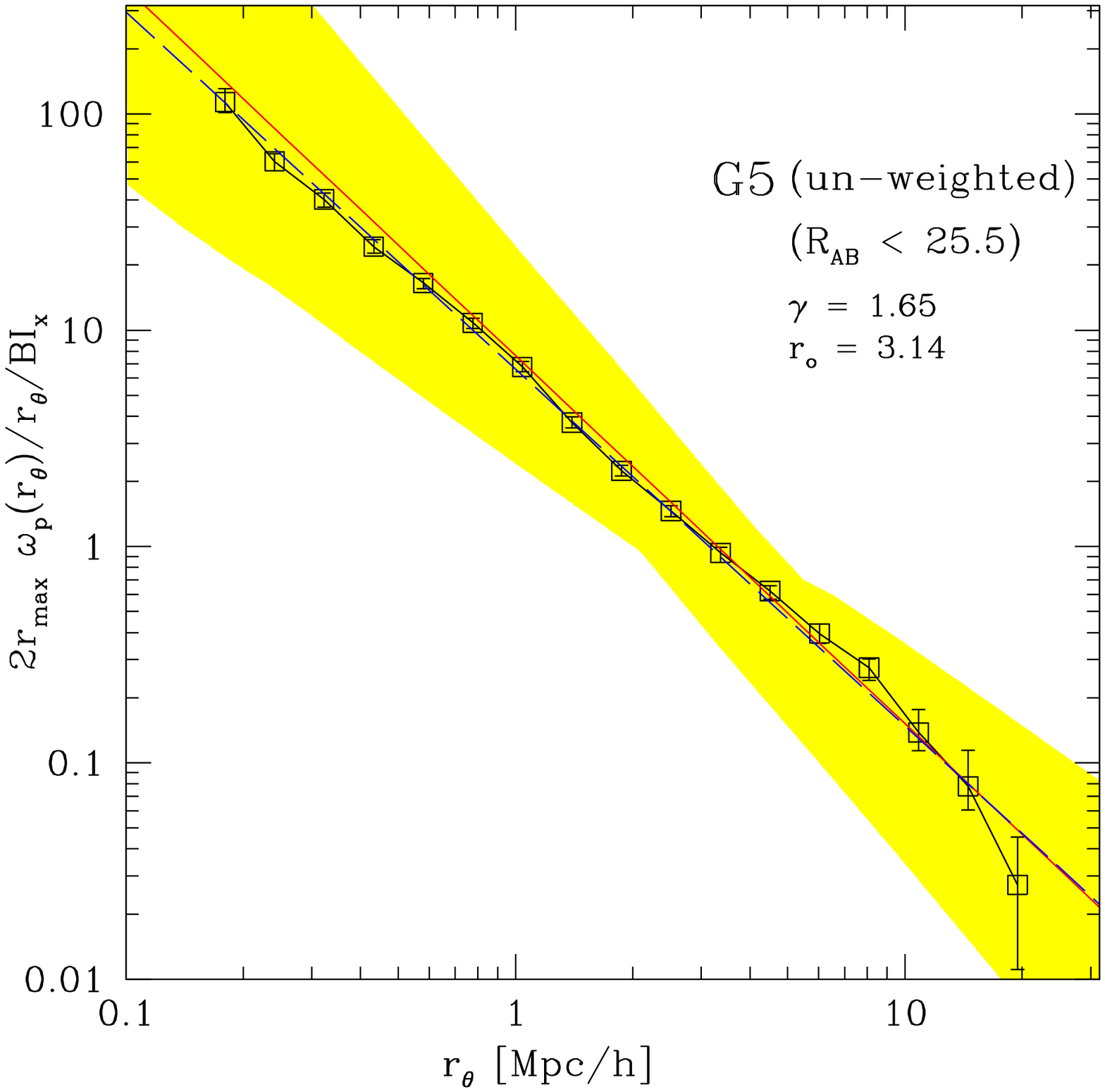}}
\caption{DLA-LBG {\it angular} CCFs at $z=3$ computed by the {\it
    unweighted} method for the D5 and G5 runs.  Other features are the
  same as described in Figure~\ref{fig:ccuw}.
\label{fig:ac}}
\end{center}
\end{figure*}

\begin{figure*}
\begin{center}
\resizebox{8.2cm}{!}{\includegraphics{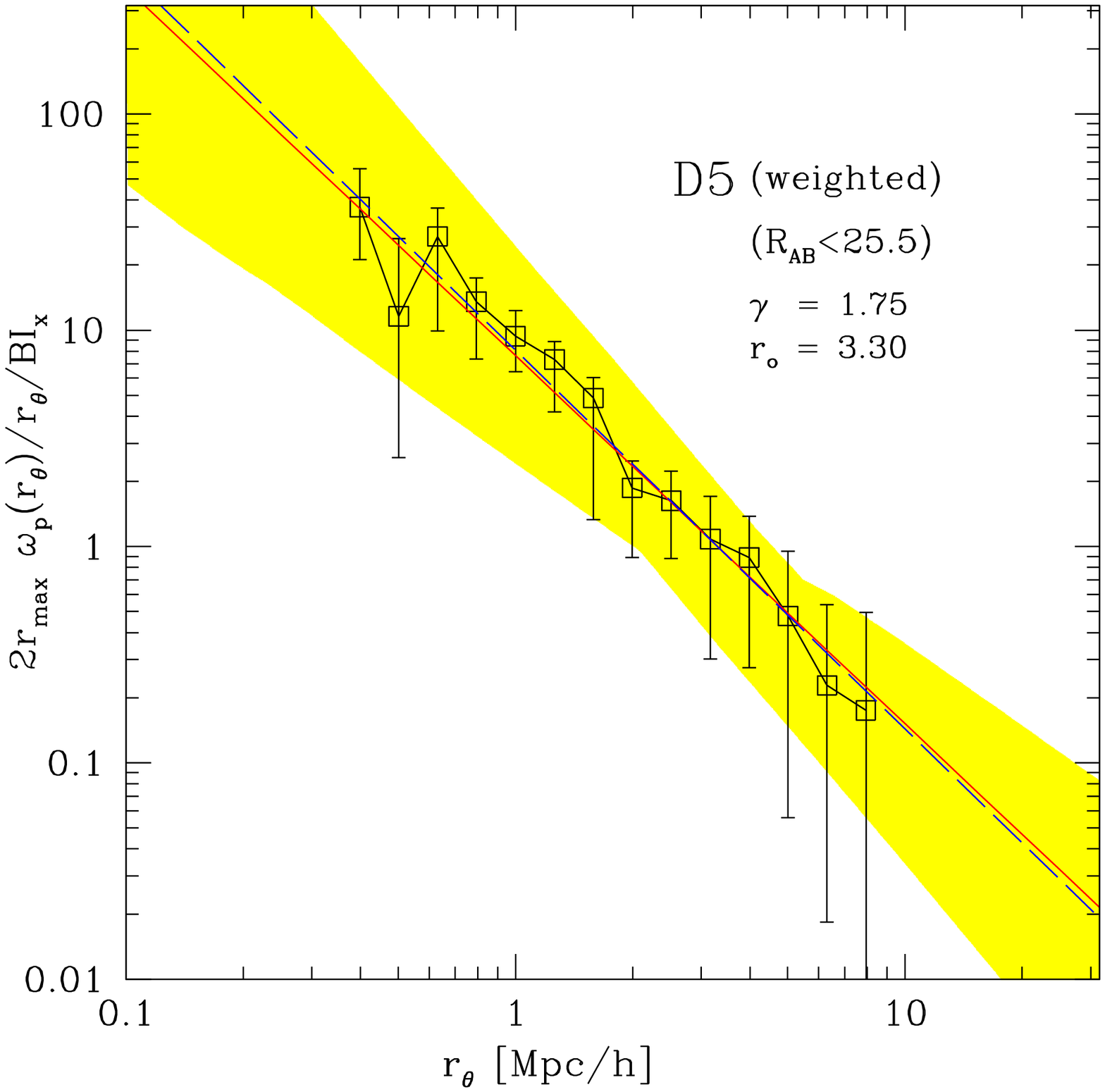}}
\hspace{0.3cm}
\resizebox{8.2cm}{!}{\includegraphics{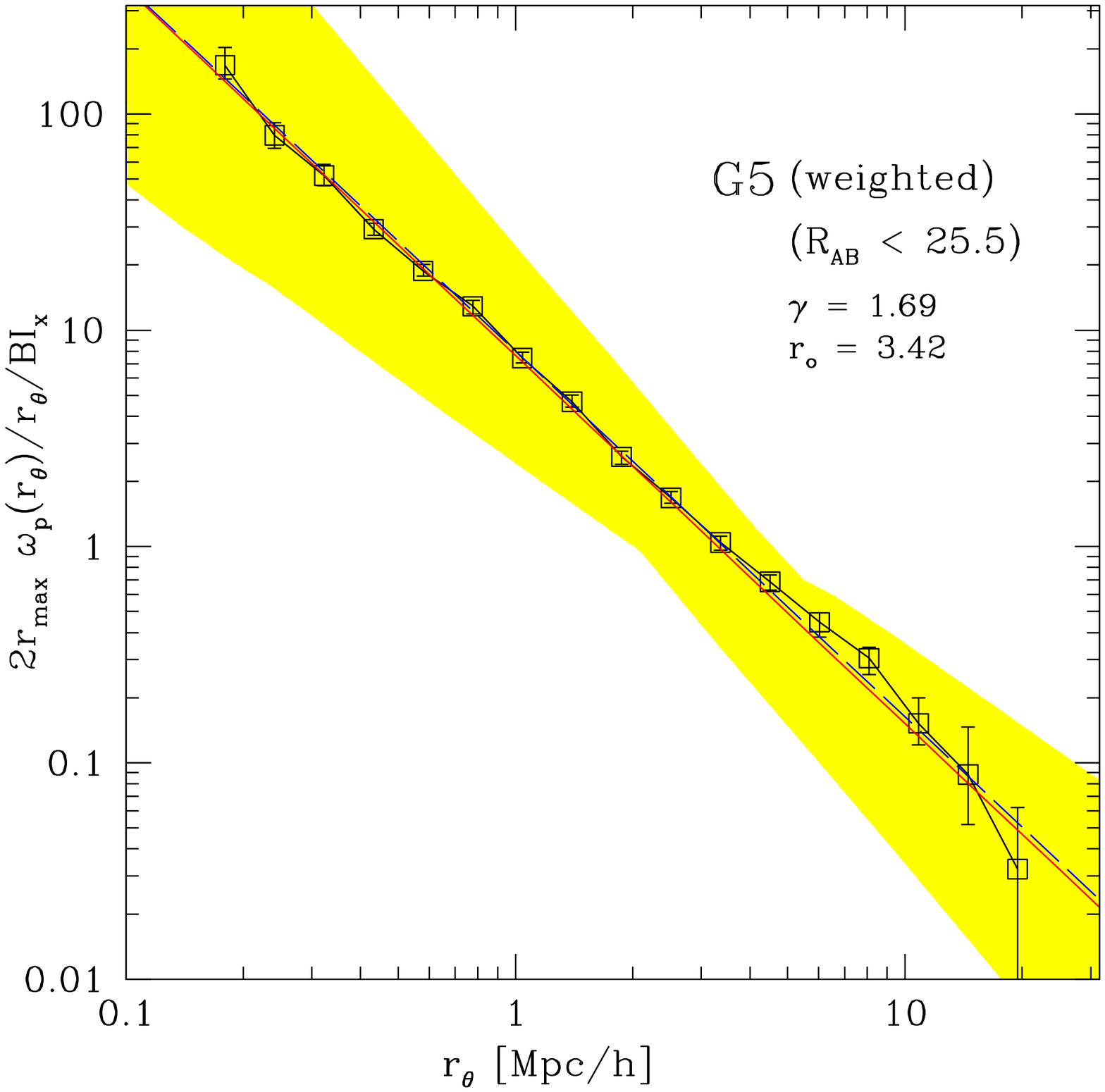}}
\caption{DLA-LBG {\it angular} CCFs at $z=3$ computed by the
  $\sDLA$-weighted method for the D5 and G5 runs.  Other features are
  the same as described in Figure~\ref{fig:ccuw}.  }
\label{fig:ac-weighted}
\end{center}
\end{figure*}

The angular CCF results of our calculations are shown in
Figures~\ref{fig:ac} and \ref{fig:ac-weighted} for both the unweighted
and the $\sDLA$-weighted method.  The best-fitting power-law
parameters are given in Table~\ref{table:acf}.  Again, the agreement
with the results of \citet{Cooke06a, Cooke06b} is within a good range.
Similarly to the 3-D CCF case, the $\sDLA$-weighted case gives a
slightly larger $\ro$ and steeper $\gamma$ than the unweighted
case.  The unweighted case of D5 is shallow with $\gamma=1.55$, but in
the $\sDLA$-weighted case, $\gamma\simeq 1.75$ is recovered.

\begin{table}
\begin{center}
\begin{tabular}{ccccccc}
\hline
& \multicolumn{2}{|c|}{Unweighted} &  \multicolumn{2}{|c|}{$\sDLA$-weighted} \\
Run &  $\ro$ & $\gamma$ &  $\ro$ & $\gamma$\\
\hline\hline
D5 &  2.75 $\pm$ 0.51 & 1.55 $\pm$ 0.20 &  3.30 $\pm$ 0.60 & 1.75 $\pm$ 0.23\cr
G5 &  3.14 $\pm$ 0.28 & 1.65 $\pm$ 0.09 &  3.42 $\pm$ 0.32 & 1.69 $\pm$ 0.10\cr
\hline
\end{tabular}
\caption{Best-fitting power-law parameters for the {\it angular} CCF
  at $z=3$.  The units of the parameters are the same as in
  Table~\ref{table:ccfit}.  The confidence limit statistics are described in Section~\ref{sec:chisq}.  For comparison, \citet{Cooke06b} reported
  $\ro = 3.32\pm 1.25\,\himpc$ and $\gamma = 1.74\pm 0.36$ for
  their angular CCF.  }
\label{table:acf}
\end{center}
\end{table}


\section{Auto-correlation functions}
\label{sec:auto}

\subsection{LBG auto-correlation}
\label{sec:LBGauto}

The auto-correlation function (ACF) also gives important constraints
on the distribution of the population under study.  In this section,
we calculate the 3-D LBG ACF by changing all subscripts in
Equation~(\ref{eq:xi01}) to `LBG':
\begin{eqnarray}
\label{eq:acf01}
\lefteqn{\xi_{\rm LBG-LBG}(r) =  {} } \nonumber\\
& & {} \frac{D_{\rm LBG} D_{\rm LBG} - 2D_{\rm LBG} R_{\rm LBG} + R_{\rm LBG} D_{\rm LBG}}{R_{\rm LBG} R_{\rm LBG}} {}. \nonumber\\
& & {}
\end{eqnarray}

\begin{figure}
\begin{center}
\resizebox{8.2cm}{!}{\includegraphics{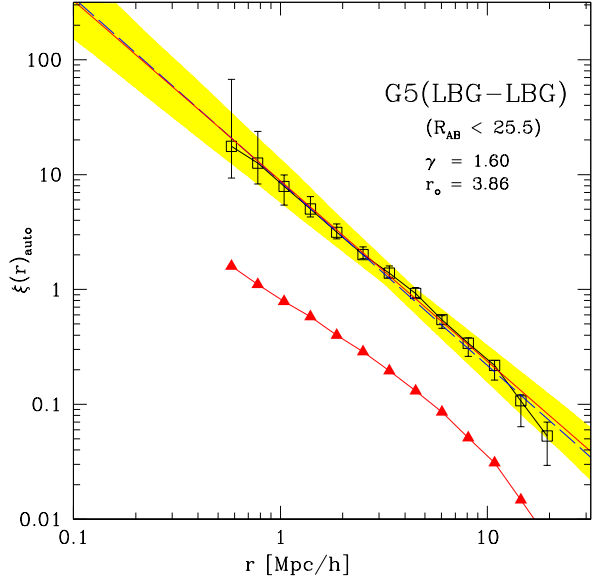}}
\caption{LBG auto-correlation function at $z=3$ for the G5 run.  The
  yellow shade shows the 1-$\sigma$ range of the best-fitting
  power-law of \citet{Ade05f}.  The variance of the ACF using 100
  random seeds is shown with vertical errorbars.  The red solid and blue
  dashed lines are the best-fitting power-laws of \citet{Ade05f} and
  this work, respectively.  The last two data points were not included
  for the power-law fit.  The red filled triangles show the dark
  matter ACF at the same redshift. }
\label{fig:LBGauto}
\end{center}
\end{figure}

Our result for the ACF is shown in Figure~\ref{fig:LBGauto}, and the
best-fitting power-law parameters (see Section~\ref{sec:chisq} for confidence limit
statistics) are $\ro = 3.86\pm0.13\,\himpc$ and
$\gamma = 1.60\pm0.07$.  The last two data points were not included for the
power-law fit because they are likely underestimated owing to the
limited box-size.  Our values of $\ro$ and $\gamma$ agree well with
the observational estimates of \citet{Ade03} and \citet{Ade05f}, who
measured the values of $\ro = 4.0 \pm 0.6\,\himpc$ and $\gamma =
1.57 \pm 0.14$ for the LBG ACF at $z \sim 3$, with a correction for the
integral constraint.  

The dark matter ACF (the red filled triangles in
Figure~\ref{fig:LBGauto}) was also computed as described in
\citet{Nag08b} in order to calculate the bias of LBGs against the
dark
matter distribution (see Section~\ref{sec:bias}).

\subsection{DLA auto-correlation}
\label{sec:DLAauto}

Similarly to the LBG ACF, it would be useful to compute the DLA ACF in
order to estimate the DLA host halo mass.  Observers also may be able to 
calculate the DLA ACF in the future when they accumulate a large enough 
sample of DLAs. In this section, we
calculate the DLA ACF with both the unweighted and the
$\sDLA$-weighted methods.  By replacing all subscripts to `DLA' in
Equations~(\ref{eq:acf01}) and (\ref{eq:xi02}), we obtain
\setlength\arraycolsep{4pt}
\begin{eqnarray}
\lefteqn{\textstyle \xi_{\rm DLA-DLA}(r) =  {} } \nonumber\\
& & {} \frac{\textstyle D_{\rm DLA}D_{\rm DLA}-2D_{\rm DLA}R_{\rm DLA}+R_{\rm DLA}R_{\rm \rm DLA}}{\textstyle R_{\rm DLA}R_{\rm DLA}} {} \nonumber\\
\label{eq:acf02a}
\end{eqnarray}
and
\setlength\arraycolsep{4pt}
\begin{eqnarray}
\lefteqn{\textstyle \xi_{\rm DLA-DLA}^{\rm weighted}(r) =  {} } \nonumber\\
& & {} \frac{\scriptstyle N_i N_j D_{\rm DLA}^i D_{\rm DLA}^j - 2 N_i N_j D_{\rm DLA}^i R_{\rm DLA}^j + N_i N_j R_{\rm DLA}^i R_{\rm DLA}^j} {\scriptstyle N_i N_j R_{\rm DLA}^i R_{\rm DLA}^j}, {} \nonumber\\
\label{eq:acf02b}
\end{eqnarray}
where $N_i N_j D_{\rm DLA}^i D_{\rm DLA}^j$ and $N_i N_j D_{\rm DLA}^i
R_{\rm DLA}^j$ are the numbers of data-data pairs and data-random
pairs, weighted by the number of DLA pixels $N_i$ and $N_j$.  As
before, 100 different realisations of random dataset have been used to
examine the statistical variance.

\begin{figure*}
\begin{center}
\resizebox{8.2cm}{!}{\includegraphics{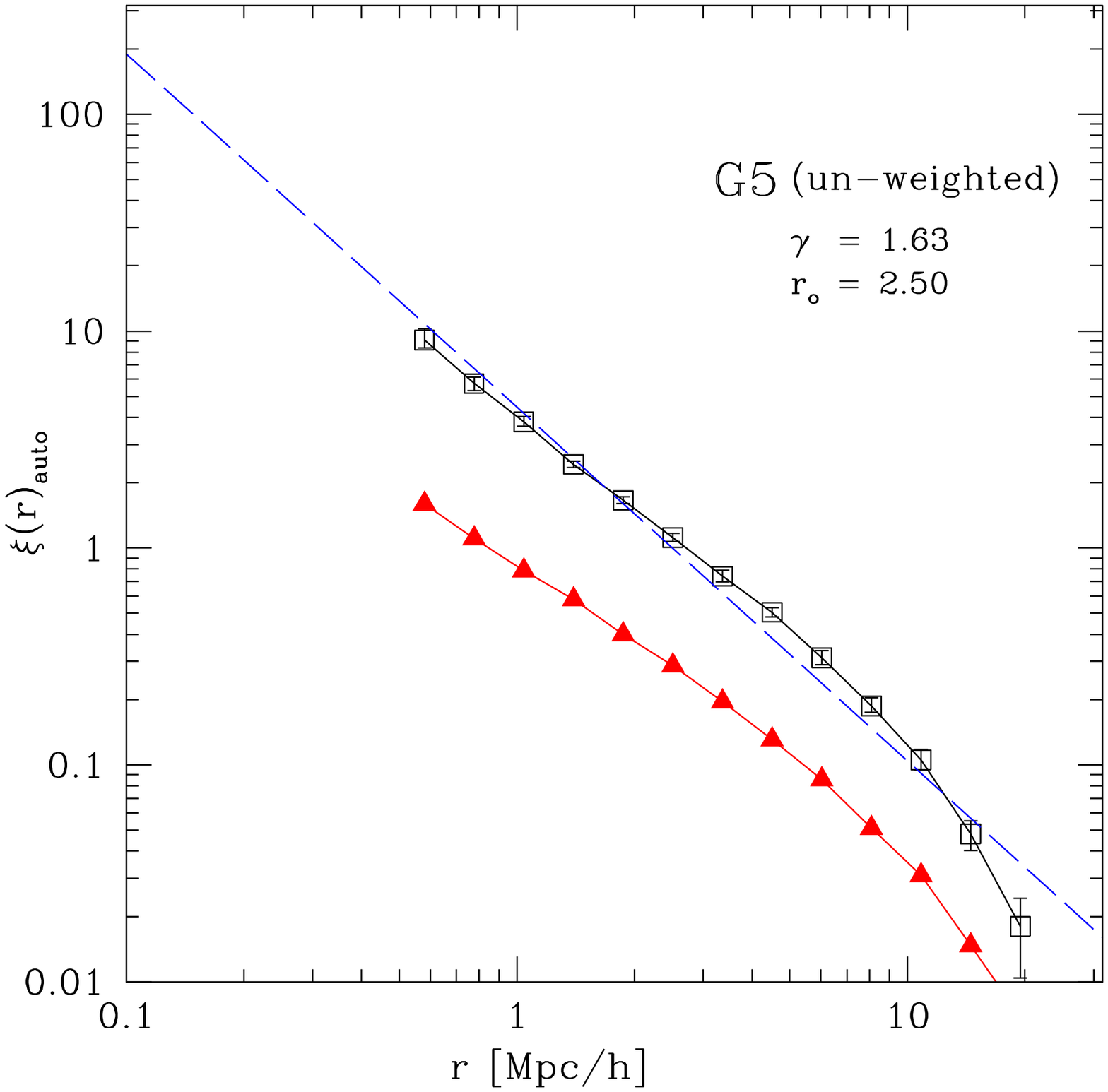}}
\hspace{0.3cm}
\resizebox{8.2cm}{!}{\includegraphics{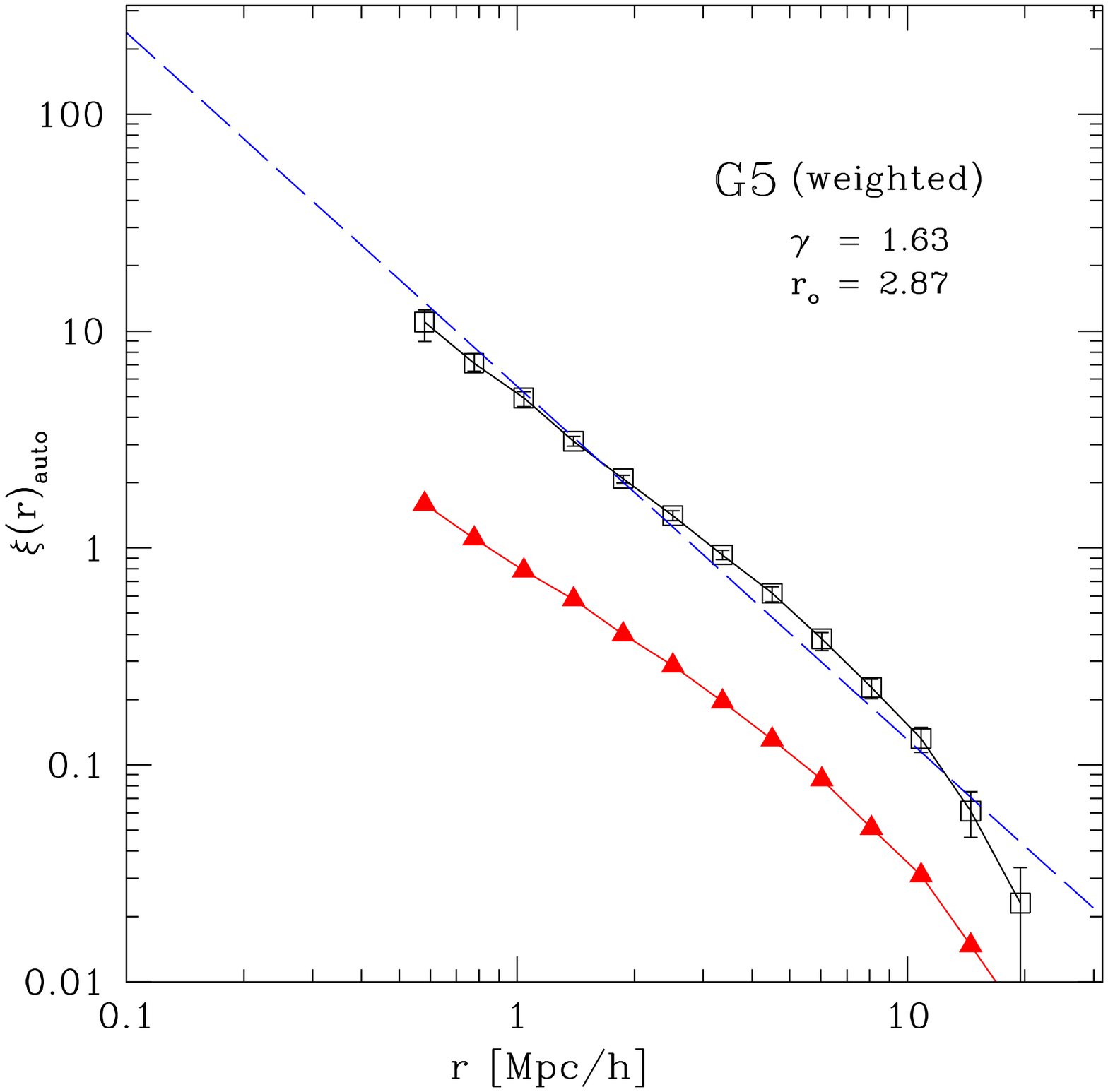}}
\caption{DLA auto-correlation function calculated with {\it
    unweighted} and {\it $\sDLA$-weighted} method for the G5 run.  The
  variance of ACFs using 100 random seeds is shown with vertical errorbars.
  The blue dashed lines are best-fits for this work. }
\label{fig:DLAauto}
\end{center}
\end{figure*}

Our DLA ACF result is shown in Figure~\ref{fig:DLAauto}, and we find
the best-fitting power-law parameters (see Section~\ref{sec:chisq} for confidence
limit statistics) of $\ro = 2.50\pm0.03\,\himpc$ and
$\gamma = 1.63\pm0.02$ for the unweighted ACF, and $\ro = 2.87\pm0.05\,\himpc$
and $\gamma = 1.63\pm0.03$ for the $\sDLA$-weighted ACF, as summarised in
Table~\ref{table:acfitdladla}.  The values of $\gamma$ are similar to
those for the LBG ACF with $\gamma \simeq 1.6$, but $\ro$ is much
smaller.  This is owing to the lower average DLA halo mass compared to
the LBG host haloes, as we will discuss further in
Section~\ref{sec:bias}.

\begin{table}
\begin{center}
\begin{tabular}{cccc}
\hline
 & & $\ro$ & $\gamma$ \\
\hline\hline
LBG-auto & & 3.86 $\pm$ 0.13 & 1.60 $\pm$ 0.07 \cr
DLA-auto (unweighted) & & 2.50 $\pm$ 0.03 & 1.63 $\pm$ 0.02 \cr
DLA-auto ($\sDLA$-weighted) & & 2.87 $\pm$ 0.05 & 1.63 $\pm$ 0.03 \cr
\hline
\end{tabular}
\caption{ACFs of LBGs and DLAs for the G5 run. The results of
  unweighted and $\sDLA$-weighted methods are given for the DLA ACF.
  $\ro$ is in units of $\himpc$.  The confidence limit statistics are described in Section~\ref{sec:chisq}.  For comparison, \citet{Ade05f}
  reported $\ro = 4.0 \pm 0.6\,\himpc$ and $\gamma = 1.57 \pm 0.14$
  for the LBGs at $z \simeq 3$.}
\label{table:acfitdladla}
\end{center}
\end{table}

\section{Bias and Halo Masses }
\label{sec:bias}

\begin{figure}
\begin{center}
\resizebox{8.2cm}{!}{\includegraphics{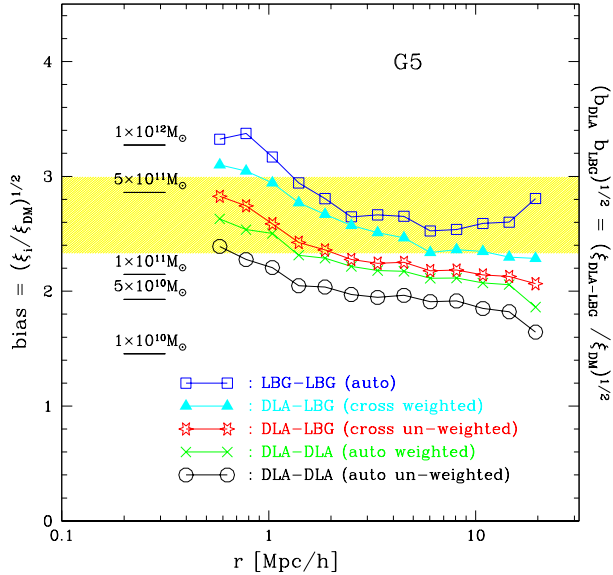}}
\caption{The biases of all correlation functions at $z=3$ that we
  computed in this paper for the G5 run.  The tick marks on the
  left-hand-side show the host halo masses calculated with the method
  described in \citet{Mo02}.  The yellow shade shows the upper and
  lower limits by \citet{Ade05f}.  }
\label{fig:bias}
\end{center}
\end{figure}

Comparing the correlation functions of DLAs and LBGs with that of dark
matter gives the measure of `bias' for the spatial distribution of
these populations against that of dark matter.  
Each observation probes certain spatial scales, and one can compute 
the average bias and a corresponding average halo mass of 
the observed sample.  In Figure~\ref{fig:bias}, we show the bias 
of the simulated DLAs and LBGs, 
defined as $b \equiv \sqrt{\xi_{i}/\xi_{\rm DM}}$, 
as a function of distance $r$, where $i={\rm LBG}$ or ${\rm DLA}$.  
This definition is based on the linear bias model,
\begin{equation}
\xi_{i}(r) = b_i^2\, \xi_{\rm DM}(r).
\end{equation}
The corresponding expression for the cross-correlation is \citep{Gawiser01}
\begin{equation}
\xi_{\rm DLA-LBG}(r) = b_{\rm DLA}\, b_{\rm LBG}\, \xi_{\rm DM}(r).
\end{equation}
Therefore, the two lines for the CCF in Figure~\ref{fig:bias} are in
fact showing $\sqrt{b_{\rm DLA}b_{\rm LBG}}$, as indicated on the axis
on the right-hand-side.  Taking the ratio of the above two expressions
gives \citep{Cooke06b}
\begin{equation}
\frac{\xi_{\rm DLA-LBG}(r)}{\xi_{\rm LBG}} = \frac{b_{\rm DLA}}{b_{\rm LBG}}.
\end{equation}

We also show the observed range of bias for the LBGs at $z=3$ by
\citet{Ade05f} as a yellow shaded region. 
In all cases shown in Figure~\ref{fig:bias}, the bias of simulated 
DLAs and LBGs slowly decreases with increasing distance.  
The upturn at $r=20\,\himpc$ for the LBG ACF is probably just noise.  
We take a simple average of bias values across the logarithmic bins 
at $r = 1.40 - 14.5\,\himpc$, and find
$\bar{b} = 2.65$, $2.48$, $2.24$, $2.17$ and $1.94$ for LBG ACF, DLA-LBG CCF
($\sDLA$-weighted), DLA-LBG CCF (unweighted), DLA ACF ($\sDLA$-weighted), 
and DLA ACF (unweighted), respectively.  The values of
$\ro$ also reflect the sizes of average bias values.  We took the
above range of scales for taking the average because most of the recent
observations are probing the scale of $r\simeq 1-10\,\himpc$.

\citet{Gawiser07} used the results of \citet{Ade05f} to obtain an
average bias of $\bar{b}_{\rm LBG}=2.5\pm 0.4$ for LBGs at $z\sim 3$.
Our average bias value of 2.60 for the LBG ACF is very close to that
of \citet{Ade05f}, and at the lower end of the estimate of
$\bar{b}_{\rm LBG} = 3.0\pm 0.5$ by \citet{LeeKS06}

The model of \citet{Sheth99} shows that an understanding of the
unconditional mass function can provide an accurate estimation of the
large-scale bias factor.  From our average bias, we calculate the mean
halo mass for LBGs and DLAs (using the unweighted and the
$\sDLA$-weighted results) based on the method described in
\citet{Mo02}, as shown in Table~\ref{table:bias}.  Our calculation of
LBG halo mass is very close to that by \citet{Ade05f}, $\Mhalo^{\rm
  LBG} = 10^{11.2} - 10^{11.8} \Msun$ (yellow shade in
Fig.~\ref{fig:bias}), which is very encouraging.  Finally,
\citet{Bouche05} estimated $\avg{\log M_{\rm DLA}} = 11.13\pm0.13$
from observations and $\avg{\log M_{\rm DLA}} = 11.16$ from
simulations.  These values are somewhat higher than the upper limit of
our unweighted DLA halo mass and close to our $\sDLA$-weighted
one.  \citet{Cooke06a} also obtained a similar value of $\Mhalo \simeq
10^{11.2} \Msun$.

Alternatively, we can directly calculate the mean DLA halo mass using
the simulation result without going through the bias argument.  For
the G5 run, the mean is $\log \avg{\Mhalo^{\rm DLA}} = 11.5$ and
$\avg{\log \Mhalo^{\rm DLA}} = 11.3$.  These values are somewhat
higher than the mean halo mass reported in Table~\ref{table:bias}.
However, the values of $\avg{\Mhalo}$ in Table~\ref{table:bias} are
computed from the average bias within the range of $r = 1.40 -
14.5\,\himpc$, and they could become higher if we included the bins at
smaller scales.  Since observers probe mostly $r\simeq 1 -
10\,\himpc$, the values reported in Table~\ref{table:bias} are more
appropriate for comparison with observations.

\citet{Bouche04} defined the parameter $\alpha$ as the ratio of
correlation functions: $\alpha\equiv\overline{b}_{\rm CCF}(M_{\rm
  DLA})/\overline{b}_{\rm ACF}(M_{\rm LBG})$.  If the value of
$\alpha$ is larger (or smaller) than unity, then the mean halo mass of
DLAs is more (or less) massive than that of the LBGs.  The ratio of
the average bias of LBG ACF and DLA-LBG CCF is $\alpha = 0.727$ for
our results.  This value is in good agreement with the observational
estimates of $\alpha = 1.62\pm 1.32$ \citep{Bouche04},
$\alpha=0.73\pm0.08$ \citep{Bouche05}, and $\alpha=0.771$
\citep{Bouche05,Mo02}.

\begin{table}
\begin{center}
\begin{tabular}{cccc}
\hline
 & & $\overline{bias}$ & $\log \avg{\Mhalo}$ \\
\hline\hline
LBG-auto & & $2.67^{+0.28}_{-0.06}$ & $11.53^{+0.22}_{-0.06}$ \cr
DLA-auto (unweighted) & & $1.94^{+0.11}_{-0.13}$ & $10.71^{+0.16}_{-0.19}$ \cr
DLA-auto ($\sDLA$-weighted) & & $2.17^{+0.14}_{-0.13}$ & $11.02^{+0.14}_{-0.16}$ \cr
\hline
\end{tabular}
\caption{Average biases and halo masses of LBGs and DLAs for the G5
  run.  The plus and minus values next to the average bias show the upper
  and lower limits at $1.40 < r < 14.5$ $\himpc$. Mean halo masses are
  computed from the second column following \citet{Mo02} and given in
  units of $\Msun$.  }
\label{table:bias}
\end{center}
\end{table}

\section{Discussion and Conclusions}
\label{sec:discussion}

Our study represents a first attempt to calculate the DLA-LBG
cross-correlation function at $z=3$ using cosmological SPH
simulations.  We calculated the DLA-LBG CCFs in several different
approaches: 3-D, angular, unweighted, and $\sDLA$-weighted.  We also
computed the auto-CF of LBGs and DLAs, and the bias against dark
matter.  In comparison to the observational data by \citet{Ade05f,
  Cooke06a, Cooke06b}, we find good agreement between our simulations
and observational measurements.  Our results suggest that the spatial
distribution of DLAs and LBGs are strongly correlated.

Let us summarise some of the main conclusions of this work. In the
first part of this paper, our results on the 3-D CCF calculated with
spherical shells (Table~\ref{table:ccfit}) are to be compared with the
3-D spherical shell result by Cooke (private communication), $\ro =
3.39 \pm 1.2 \,\himpc$ and $\gamma = 1.61 \pm 0.3$.  Our results are
consistent with Cooke's within the error.  The shallow slope of
Cooke's above estimate probably owes to the limited sample size in
the spherical shell at small distances, as we discussed in
Sections~\ref{sec:ccf} and \ref{sec:angular}.

In the second part, we have replaced the spherical shell method with
the projected approach used in \citet{Ade03} and \citet{Cooke06b}, and
calculate the best-fitting values given in Table~\ref{table:acf}.
Encouragingly, our results are within the upper and lower limits of
the observational measurement by \citet{Cooke06a,Cooke06b}. We corrected
all CFs in this paper with the integral constraint.

Finally, we also analysed the auto-correlation functions of LBGs and
DLAs at $z=3$ (Table~\ref{table:acfitdladla}) found in our
simulations. Our results for the best-fitting parameters of the LBG
ACF agree well with \citet{Ade05f}. Our results show
that LBGs are more strongly correlated than DLAs, and have higher mean
halo mass.

\begin{figure}
\begin{center}
\resizebox{8.2cm}{!}{\includegraphics{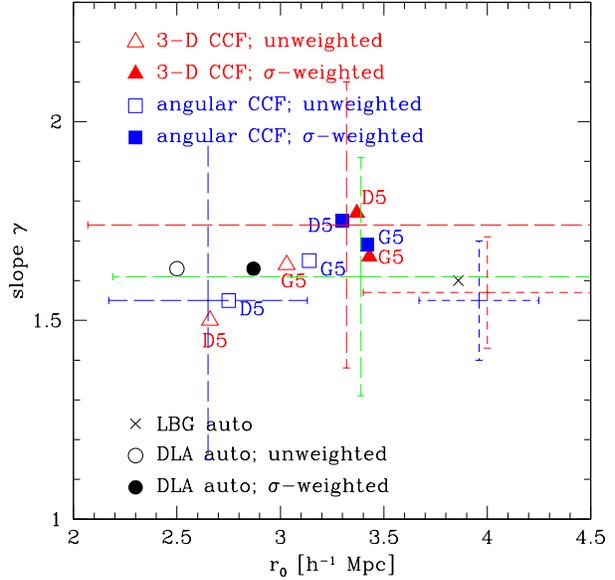}}
\caption{Summary of best-fitting power-law parameters for all
  correlation functions that we obtained in earlier sections. Long blue, 
  red, and green dashed cross lines are for the LBG ACF, the angular CCF, and the 3-D CCF 
  of \citet{Cooke06a, Cooke06b}, respectively. The LBG ACF of \citet{Ade03} 
  is shown in a short blue dashed line and of \citet{Ade05f} is shown in red.}
\label{fig:param}
\end{center}
\end{figure}

Figure~\ref{fig:param} summarises the best-fitting power-law
parameters for all the correlation functions that we obtained in the
earlier sections.  In most cases, the slope $\gamma$ falls into the
range $\approx 1.5-1.7$ and the variation is not very large. The
correlation length $\ro$ shows a larger variation from
$2.5\,\himpc$ to $4\,\himpc$, depending on the sample and calculation
method.  This trend is similar to that seen by
\citet[][Fig.\,8]{Cooke06b}.  In general, the $\sDLA$-weighted method
gives a larger $\ro$ than the unweighted method.


Finally, the LBG bias, derived from the LBG ACF in
Section~\ref{sec:bias}, has led to the upper and lower limits of the
LBG dark matter halo mass of $\log \avg{\Mhalo} =
11.53^{+0.22}_{-0.06}$ (see Table~\ref{table:bias}).  This result is
consistent with observational estimates of the LBG halo mass of
$\Mhalo\sim 10^{12} \Msun$, \citep[e.g.,][]{Steidel98, Ade98} and
within the limit of $\Mhalo=10^{11.2} - 10^{11.8}M_{\odot}$
\citep{Ade05f}. Similarly, we derived the DLA biases, and obtained the
mean DLA halo masses as shown in
Table~\ref{table:bias}. \citet{Cooke06a}'s measurement showed a DLA
galaxy bias of $b_{\rm DLA}\sim 2.4$ and an average DLA halo mass of
$\Mhalo \sim 10^{11.2} \Msun$.  Our average DLA bias ($\overline{b}=1.94$ and 
$\overline{b}=2.17$ for un-weighted DLA ACF and weighted DLA ACF, respectively) and halo mass
estimates ($\log\avg{\Mhalo^{\rm DLA}}$=10.71 and 11.02 for un-weighted DLA 
ACF and weighted DLA ACF, respectively) are in good agreement with theirs.  
We also examined the ratio of bias values defined as $\alpha\equiv\overline{b}_{\rm 
CCF}/\overline{b}_{\rm ACF}$ \citep{Bouche04}, and found that our
value of $\alpha = 0.727$ agrees well with the observational
estimates.  This again shows that the mean halo mass of DLAs is less
than that of the LBGs. The fact that $\avg{\Mhalo^{\rm LBG}}$ is greater than
$\avg{\Mhalo^{\rm DLA}}$ is a natural outcome because the LBG sample
is limited to the bright star-forming galaxies with $R_{\rm AB} < 25$
and $M_{\star} \simeq 10^{10}-10^{11} \Msun$, whereas the DLA \HI\ gas
is present in numerous lower mass haloes below the LBG threshold. 

Motivated by our earlier successes of our simulations to reproduce 
the physical properties of LBGs such as stellar mass, SFR, and 
colours \citep{Nag04e}, in this work we examined another observational 
method, i.e. DLA-LBG CCF, to further check the consistency 
between observations and simulations. We found good agreement between our
results and observations. Furthermore, there are accumulated evidence 
that suggest high halo masses for LBGs \citep[e.g.,][]{ Mo96, Ade98,Baugh98, 
Giavalisco98, Steidel98, Kau99, Mo99, Katz99, Papovich01, Shapley01}. 
Therefore, the scenario that the majority of LBGs is merger-induced 
starburst systems associated with low-mass haloes 
\citep{Lowenthal97, Sawicki98,Som01,Wea03} does not appear to be 
a favored model for LBGs.

In our simulations, we estimated the \HI\ column densities using a
pixel size that is much larger than the typical quasar beam size,
which is of the order of parsecs. This may have some impact on our
estimates of $\NHI$ and the corresponding statistics such as the \HI\ 
column density distribution function. For example, if the ISM is
clumpy on smaller scales than our pixel size, there could be
high-density neutral clouds below our resolution scale that are
self-shielded and contain larger amounts of \HI.  Unfortunately, owing
to limitations in computational resources, it is not possible for us
at the moment to run such a high-resolution cosmological simulation
with the same box size as we have used in this paper.  In future work,
we will nevertheless attempt to check the dependence of our $\NHI$
estimates on numerical resolution, and perform more rigorous
resolution tests.


\section*{Acknowledgments}

This work is supported in part by the National Aeronautics and Space
Administration under Grant/Cooperative Agreement No. NNX08AE57A issued
by the Nevada NASA EPSCoR program, and the President's Infrastructure
Award at UNLV.  KN is grateful for the hospitality of the Institute for
the Physics and Mathematics of the Universe (IPMU), University of Tokyo,
where part of this work was done.  The simulations were performed
at the Institute of Theory and Computation at Harvard-Smithsonian
Center for Astrophysics, and the analysis were performed at the UNLV
Cosmology Computing Cluster.



\bsp

\label{lastpage}

\end{document}